\documentclass[aps, prd, letterpaper, amsmath, amssymb, superscriptaddress, twocolumn, floatfix, nofootinbib, longbibliography]{revtex4-1}
\usepackage[pdftex]{graphicx}
\usepackage[pdftex, pdfstartview={FitH}, pdfnewwindow=true, colorlinks=true, citecolor=blue, filecolor=blue, linkcolor=blue, urlcolor=blue, pdfpagemode=UseNone, bookmarks=false]{hyperref}
\usepackage{placeins} 
\usepackage{comment}
\usepackage{wrapfig}
\usepackage[T1]{fontenc} 
\usepackage{verbatim}
\usepackage[export]{adjustbox}
\usepackage{amssymb}
\usepackage{dsfont}
\usepackage{lipsum}

\newcommand{\Ibb}{\ensuremath{\mathbb I} }

\newcommand{\Tr}{\ensuremath{\mbox{Tr}\;} }
\newcommand{\vev}[1]{\ensuremath{\left\langle #1 \right\rangle} }
\newcommand{\nn}{\nonumber}

\begin{document}
\title{\boldmath $SO(4)$ invariant Higgs-Yukawa model with reduced staggered fermions }

\author{Nouman Butt}
\affiliation{Department of Physics, Syracuse University, Syracuse, New York 13244, United States}
\author{Simon Catterall}\email{smcatter@syr.edu}
\affiliation{Department of Physics, Syracuse University, Syracuse, New York 13244, United States}
\author{David Schaich}
\affiliation{AEC Institute for Theoretical Physics, University of Bern, 3012 Bern, Switzerland}
\date{14 October 2018}
\begin{abstract}
We explore the phase structure of a four dimensional $SO(4)$ invariant lattice Higgs-Yukawa model comprising four reduced staggered fermions interacting with a real scalar field.
The fermions belong to the fundamental
representation of the symmetry group while the three scalar field components transform in the self-dual representation of $SO(4)$. The model is a generalization of a four fermion system with the same
symmetries that has received recent attention because of its unusual phase structure comprising massless and massive symmetric phases separated by a very narrow phase in which a small
bilinear condensate breaking $SO(4)$ symmetry is present. The generalization described in this paper simply consists of the addition of a scalar kinetic term. We find a region of the enlarged phase diagram
which shows no sign of a fermion condensate or symmetry breaking but in which there is nevertheless evidence of a diverging correlation length.  Our results in this region are consistent with the presence of
a single continuous phase transition separating the massless and massive symmetric phases observed in the earlier work.
\end{abstract}

\maketitle
\flushbottom

\section{Introduction}
The motivation for this work comes from recent numerical studies~\cite{Ayyar:2014eua, Ayyar:2015lrd, Catterall:2015zua, He:2016sbs, Ayyar:2016lxq, Ayyar:2016nqh, Catterall:2016dzf, Schaich:2017czc} of a particular lattice four fermion theory constructed using reduced staggered fermions~\cite{Bock:1992yr}. In three dimensions this theory appears to
exist in two phases - a free massless phase and a phase in which the fermions acquire a mass~\cite{Ayyar:2014eua, Ayyar:2015lrd, Catterall:2015zua, He:2016sbs}. What is unusual about this is that no local order parameter has been identified which distinguishes between
these two phases - the massive phase \textit{does not} correspond to a phase of broken symmetry as would be expected in a conventional Nambu--Jona-Lasinio scenario. Furthermore, the transition between these
two phases is continuous but is not characterized by Heisenberg critical exponents.

When this theory is lifted to four dimensions, however, a very narrow symmetry broken phase reappears characterized by a small bilinear condensate~\cite{Ayyar:2016lxq, Ayyar:2016nqh, Catterall:2016dzf, Schaich:2017czc}.
In Ref.~\cite{Catterall:2017ogi} two of us constructed a continuum realization of this lattice theory and argued that topological
defects may play an important role in determining the phase structure. This calculation suggests that the addition of a kinetic term for the auxiliary scalar field $\sigma^+$ used to generate the four
fermion interaction may allow access to a single phase transition between massless (paramagnetic weak-coupling, PMW) and massive (paramagnetic strong-coupling, PMS) symmetric phases.
In this paper we provide evidence in favor of this from direct numerical investigation of the lattice Higgs-Yukawa model.
This development presents the possibility of new critical behavior in a four-dimensional lattice theory of strongly interacting fermions, which would be very interesting from both theoretical and phenomenological viewpoints, and also connects to recent activity within the condensed matter community~\cite{Fidkowski:2009dba, Morimoto:2015lua}.

The plan of the paper is as follows: in the next section we describe the action and symmetries of the lattice theory, followed by a discussion of analytical results in certain limits in Sec.~\ref{sec:analytic}.
We present numerical results for the phase structure of the theory in Sec.~\ref{sec:phases}, and extend this investigation in Sec.~\ref{sec:bilin} by adding symmetry-breaking source terms to the action in order to search for spontaneous symmetry breaking in the thermodynamic limit.
These investigations reveal significant sensitivity to the hopping parameter $\kappa$ in the scalar kinetic term, with an antiferromagnetic (AFM) phase separating the PMW and PMS phases for $\kappa \leq 0$ but an apparently direct and continuous transition between the PMW and PMS phases for a range of positive $\kappa_1 < \kappa < \kappa_2$.
Our current work constrains $0 < \kappa_1 < 0.05$ and $0.085 < \kappa_2 < 0.125$.
We collect these results to present our overall picture for the phase diagram of the theory in Sec.~\ref{sec:diagram}.
We conclude in Sec.~\ref{sec:conc} by summarizing our findings and outlining future work.

\section{Action and Symmetries}
The action we consider takes the form
\begin{equation}
\begin{split}
S = \sum_{x} \psi^a [ \eta.\Delta^{ab} + G\sigma^{+}_{ab} ] \psi^{b} + \frac{1}{4}\sum_{x} (\sigma^{+}_{ab})^2 \\
- \frac{\kappa}{4} \sum_{x,\mu} \left[ \sigma^{+}_{ab}(x) \sigma^{+}_{ab}(x + \mu) + \sigma^{+}_{ab}(x) \sigma^{+}_{ab}(x-\mu) \right]
\label{eq:S}
\end{split}
\end{equation}
where repeated indices are to be contracted and $\eta^{\mu}(x)=\left(-1\right)^{\sum_{i=1}^{\mu-1}x_i}$ are the usual staggered fermion phases.
The discrete derivative is given by
\begin{equation}
\Delta_{\mu}^{ab} \psi^{b} = \frac{1}{2}\delta^{ab} [\psi^{b}(x +\mu) - \psi^{b}(x-\mu)].
\end{equation}
The self-dual scalar field $\sigma^{+}_{ab}$ is defined as
\begin{equation}
\sigma^{+}_{ab} = P^{+}_{abcd} \sigma_{cd} = \frac{1}{2} \left[ \sigma_{ab} + \frac{1}{2} \epsilon_{abcd} \sigma_{cd}\right]
\end{equation}
with $P^{+}$ projecting the antisymmetric matrix field $\sigma(x)$ to its self-dual component.

The second line in eqn.~\ref{eq:S} is essentially a kinetic operator for the $\sigma^{+}$ field. With $\kappa$ set equal to zero we can integrate out the auxiliary field and recover the pure four fermion
model studied in Ref.~\cite{Catterall:2016dzf}.  The rationale for including such a bare kinetic term for the auxiliary field is provided by arguments set out for a related
continuum model in Ref.~\cite{Catterall:2017ogi}. More concretely, it should
be clear that $\kappa>0$ favors ferromagnetic ordering of the scalar field and associated fermion bilinear. This is to be contrasted
with the preferred antiferromagnetic ordering observed in Refs.~\cite{Ayyar:2016lxq, Schaich:2017czc} for the $\kappa=0$ theory.\footnote{Although Ref.~\cite{Catterall:2016dzf} observed a strong response to an antiferromagnetic external source, evidence of spontaneous ordering in the zero-source thermodynamic limit was not found until the follow-up Ref.~\cite{Schaich:2017czc}.}
The competition between these two effects raises the
possibility that the $\kappa=0$ antiferromagnetic fermion bilinear condensate may be suppressed as $\kappa$ is increased.

In contrast to similar models studied by Refs.~\cite{Stephenson:1988td, Hasenfratz:1988vc, Lee:1989xq, Lee:1989mi, Bock:1990cx, Abada:1990ds, Hasenfratz:1991it, Gerhold:2007yb, Gerhold:2007gx} we fix the coefficient of the $((\sigma^{+})^2 - 1)^2$ term in the action to be $\lambda = 0$.
Without this term to provide a constraint on the magnitude of the scalar field, we will encounter instabilities when the magnitude of $\kappa$ is too large.
We discuss these instabilities in more detail in the next section.

In addition to the manifest $SO(4)$ symmetry the action is also invariant under a shift symmetry
\begin{equation}
\psi(x) \to \xi_{\rho} \psi(x + \rho)
\end{equation}
with $\xi_\mu(x)=\left(-1\right)^{\sum_{i=\mu}^dx_i}$
and a discrete $Z_2$ symmetry:
\begin{align}
\sigma^{+}& \to - \sigma^{+} \\
\psi^a &\to i\epsilon(x) \psi^a.
\label{eq:Z_2}
\end{align}
Both the $Z_2$ and $SO(4)$ symmetries prohibit local bilinear fermion mass terms from appearing as a result of quantum corrections.
Non-local $SO(4)$-symmetric bilinear terms can be constructed by coupling fields at different sites in the unit hypercube but such terms break the shift symmetry. Further discussion of possible bilinear mass terms is presented in detail in Ref.~\cite{Catterall:2016dzf}.

\section{\label{sec:analytic}Analytical results}
Before we present numerical results we can analyze the model in certain limits.
For example, since the action is quadratic in $\sigma^+$ we can consider the effective action obtained by integrating over $\sigma^+$. The scalar part of the
action may be rewritten
\begin{equation}
  \label{sigmaeq}
  \frac{1}{4} \sigma^+\left(-\kappa\Box+m^2\right)\sigma^+
\end{equation}
where $m^2=\left(1-2d\kappa\right)$ is an effective mass squared for the $\sigma^+$ field in $d$ dimensions and $\Box$ is the usual discrete scalar laplacian.
Integrating out $\sigma^+$ yields an effective action for the fermions
\begin{equation}
S=\sum\psi \left(\eta .\Delta\right)\psi-G^2\sum \Sigma^+\left[-\kappa\Box+m^2\right]^{-1}\Sigma^+
\end{equation}
where $\Sigma^{+}_{ab}=\left[\psi_a\psi_b\right]_+$ is the self-dual fermion bilinear.
For $\kappa$ small we can expand the inverse operator in powers of $\kappa/m^2$ and find
\begin{equation}
  S=\sum\psi \left(\eta .\Delta\right)\psi-\frac{G^2}{m^2}\Sigma^+\left(\Ibb + \frac{\kappa}{m^2}\Box+\ldots\right)\Sigma^+
  \label{effS}.
\end{equation}
To leading order the effect of non-zero $\kappa$ is to renormalize the Yukawa coupling $G\to \frac{G}{m}=\frac{G}{\sqrt{1-2\kappa d}}$.
At next to leading order we obtain the term
\begin{equation}
  \frac{G^2}{m^4} \sum \Sigma^+ \left[-\kappa \Box\right] \Sigma^+.
\end{equation}
For $\kappa>0$ and sufficiently large $G$ this term favors a ferromagnetic ordering of the fermion bilinear $\vev{\Sigma^+}\ne 0$.
Conversely it suggests an antiferromagnetic ordering with $\vev{\epsilon(x)\Sigma^+(x)}\ne 0$ for $\kappa<0$.
This can be seen more clearly if one rewrites the action in the alternative form
\begin{equation}
  S = \sum \psi \left(\eta.\Delta\right)\psi - G^2 \sum \Sigma^+ \left[-\kappa B + \Ibb\right]^{-1} \Sigma^+
\end{equation}
where $B\Sigma = \sum_{\mu} \left[\Sigma(x + \mu) + \Sigma(x - \mu)\right]$.
Clearly changing the sign of $\kappa$ can be compensated by transforming $\Sigma^+ \to \epsilon(x)\Sigma^+$ since $\epsilon(x)$ anticommutes with $B$.
Two of us investigated the case $\kappa=0$ in Ref.~\cite{Schaich:2017czc} and observed a narrow phase with antiferromagnetic ordering.
Since $\kappa>0$ produces ferromagnetic terms we expect the tendency toward antiferromagnetic ordering to be reduced as $\kappa$ is increased.
The numerical results described in the following section confirm this.

For $\kappa>\frac{1}{2d}=\frac{1}{8}$ the squared mass changes sign and one expects an instability
to set in with the model only being well defined for $\kappa<\frac{1}{8}$. Actually there is also a lower bound on
the allowed values of $\kappa$. To see this return to eqn.~\ref{sigmaeq} and perform the change of variables
\begin{align}
  \label{ktominusk}
  \sigma^{+}_{ab}(x) & \to \epsilon(x) \sigma^{+}_{ab}(x) \\
  \kappa & \to -\kappa. \nn
\end{align}
This implies that the partition function $Z\left(\kappa\right)$ is an even function of $\kappa$ at $G=0$.  We can show that
this is also true in the strong coupling limit $G\to\infty$. In this limit we can drop the fermion kinetic term
from the action in eqn.~\ref{eq:S} and expand the Yukawa term in powers of the fermion field
\begin{equation}
Z=\int D\psi D\sigma^+\left(1-G\psi\sigma^+\psi+\frac{1}{2}(G\psi\sigma^+\psi)^2\right)e^{S(\sigma^+)}.
\end{equation}
The only terms that survive the Grassmann integrations contain even powers of $\sigma^+$. Using the same transformation eqn.~\ref{ktominusk} allows us to show that the partition function is once again an even
function of $\kappa$. Thus we expect that at least for weak and strong coupling the partition function is only
well defined in the strip $-\frac{1}{8}<\kappa<\frac{1}{8}$.

It is also instructive to compute the effective action for the scalar fields having integrated out the fermions.
This takes the form\footnote{To facilitate the computation we have traded the original ferromagnetic Yukawa coupling $\psi\sigma^+\psi$ in eqn.~\ref{eq:S} for an antiferromagnetic coupling $\epsilon(x)\psi\sigma^+\psi$
while simultaneously trading $\kappa\to-\kappa$ as in eqn.~\ref{ktominusk}.
This allows us to simplify the expression for the effective action by using the fact that $\epsilon(x)$ anticommutes with $\Delta_\mu$.}
\begin{equation}
\label{eq:S-eff}
S_{\text{eff}} = -\frac{1}{4} \Tr \ln \left(-\Delta_\mu^2+M^2+G\eta_\mu(x)\epsilon(x)\Delta_\mu\sigma^+\right)
\end{equation}
where $M^2=-G^2(\sigma^+)^2$.
To zeroth order in derivatives the resultant effective potential is clearly of symmetry breaking form.
The first non-trivial term in the derivative or large mass $M$ expansion of this action is
\begin{equation}
-\frac{G^2}{8M^4}\sum \left(\Delta_\mu\sigma^+\right)^2.
\end{equation}
Thus even the pure four fermion model will produce kinetic terms for the scalar field through loop effects confirming the need to include such terms in the classical action.\footnote{A similar argument suggests that a quartic term $\lambda((\sigma^+)^2 - 1)^2$ will also be produced.  As mentioned in the previous section we fix $\lambda = 0$ in the calculations reported here.  In addition to simplifying the parameter space to be considered, this step is also motivated by observations~\cite{Gerhold:2007yb, Gerhold:2007gx} that $\lambda$ seems to have little effect on the large-scale features of the phase diagram in similar Higgs-Yukawa models.}
In Ref.~\cite{Catterall:2017ogi} it was argued that an additional term
should also be generated which is quartic in derivatives in the continuum limit. This term \textit{only} arises for
a self-dual scalar field and leads to the possibility that topological field configurations called Hopf defects
may play a role in understanding the massive symmetric phase.

\section{\label{sec:phases}Phase Structure}
\begin{figure}[tbp]
  \centering
  \includegraphics[width=0.48\textwidth]{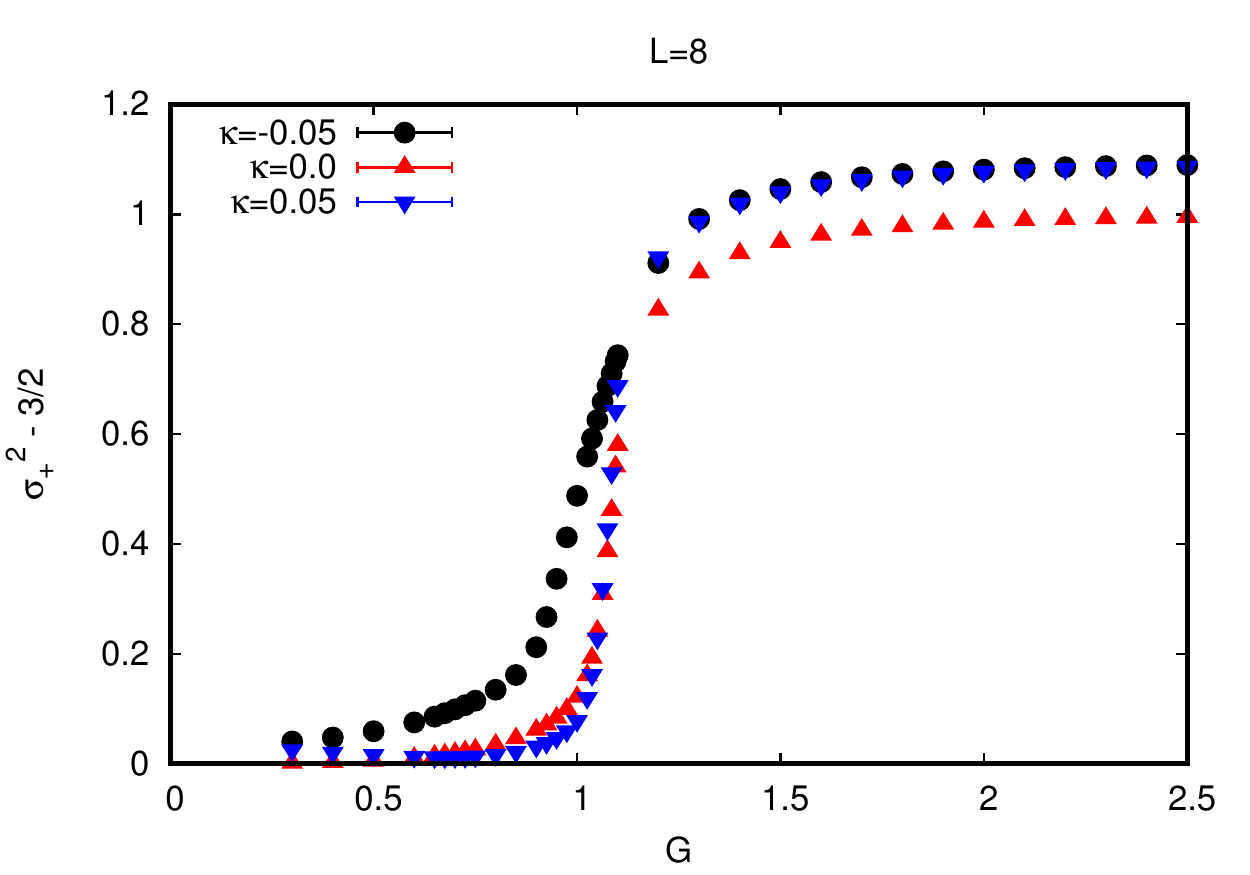}
  \caption{$\vev{\sigma^2_{+}}$ vs $G$ for $L = 8$, comparing $\kappa = \pm0.05$ and 0.}
  \label{fig:plus}
\end{figure}

One useful observable we can use to probe the phase structure in the $\left(\kappa, G \right)$ plane is
$\vev{\sigma_{+}^2}$. This is shown for
three different values of $\kappa$ on a $8^4$ lattice in Fig.~\ref{fig:plus}.
At $\kappa=0$ this observable served as a proxy for the four fermion condensate and we observe this to be
the case also when $\kappa\ne 0$. Thus we see that a four fermion phase survives at strong Yukawa coupling $G$
even for non-zero values of $\kappa$.

\begin{figure}[tbp]
  \centering
  \includegraphics[width=0.48\textwidth]{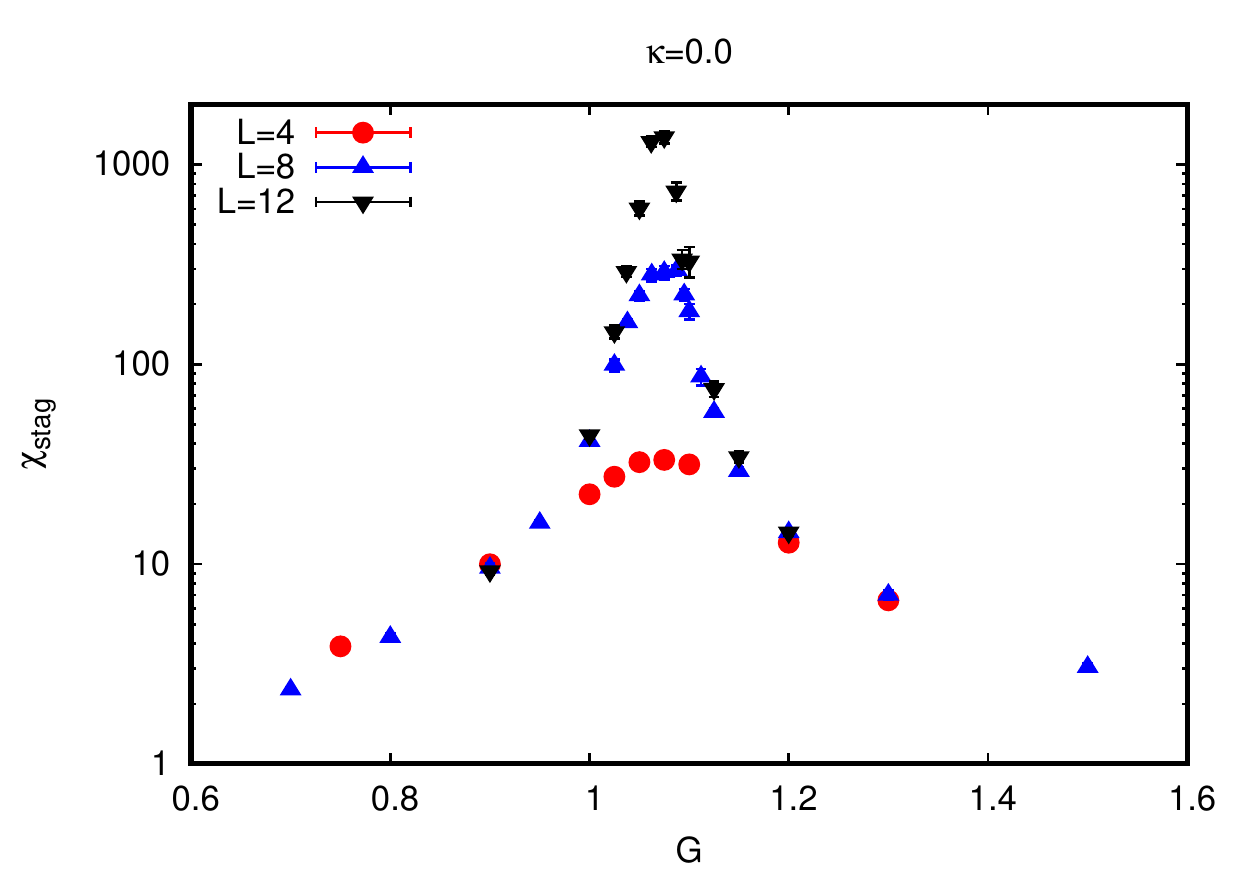}
  \caption{$\chi_{\text{stag}}$ vs $G$ at $\kappa=0$ for $L=4$, 8 and 12.} \label{fig:sus_0.0}
\end{figure}

Of course the key issue is what
happens for intermediate values of $G$. At $\kappa=0$ a narrow intermediate phase was observed for $0.95 \lesssim G \lesssim 1.15$ in two different ways: from the
volume scaling of a certain susceptibility~\cite{Ayyar:2016lxq} and by examining fermion bilinear condensates as functions
of external symmetry breaking sources~\cite{Schaich:2017czc}. This susceptibility is defined as
\begin{equation}
  \chi_{\text{stag}} = \frac{1}{V} \sum_{x,y,a,b} \vev{\epsilon(x) \psi^a(x)\psi^b(x) \epsilon(y) \psi^a(y)\psi^b(y)}
\end{equation}
where $V = L^4$ and the subscript ``$_{\text{stag}}$'' refers to the presence of the parity factors $\epsilon(x)$ associated with antiferromagnetic
ordering. It is shown in Fig.~\ref{fig:sus_0.0} for three different lattice volumes at $\kappa = 0$. The linear dependence
of the peak height on the lattice volume is consistent with the presence of a condensate $\vev{\epsilon(x)\psi^a(x)\psi^b(x)} \ne 0$.

\begin{figure}[tbp]
  \centering
  \includegraphics[width=0.48\textwidth]{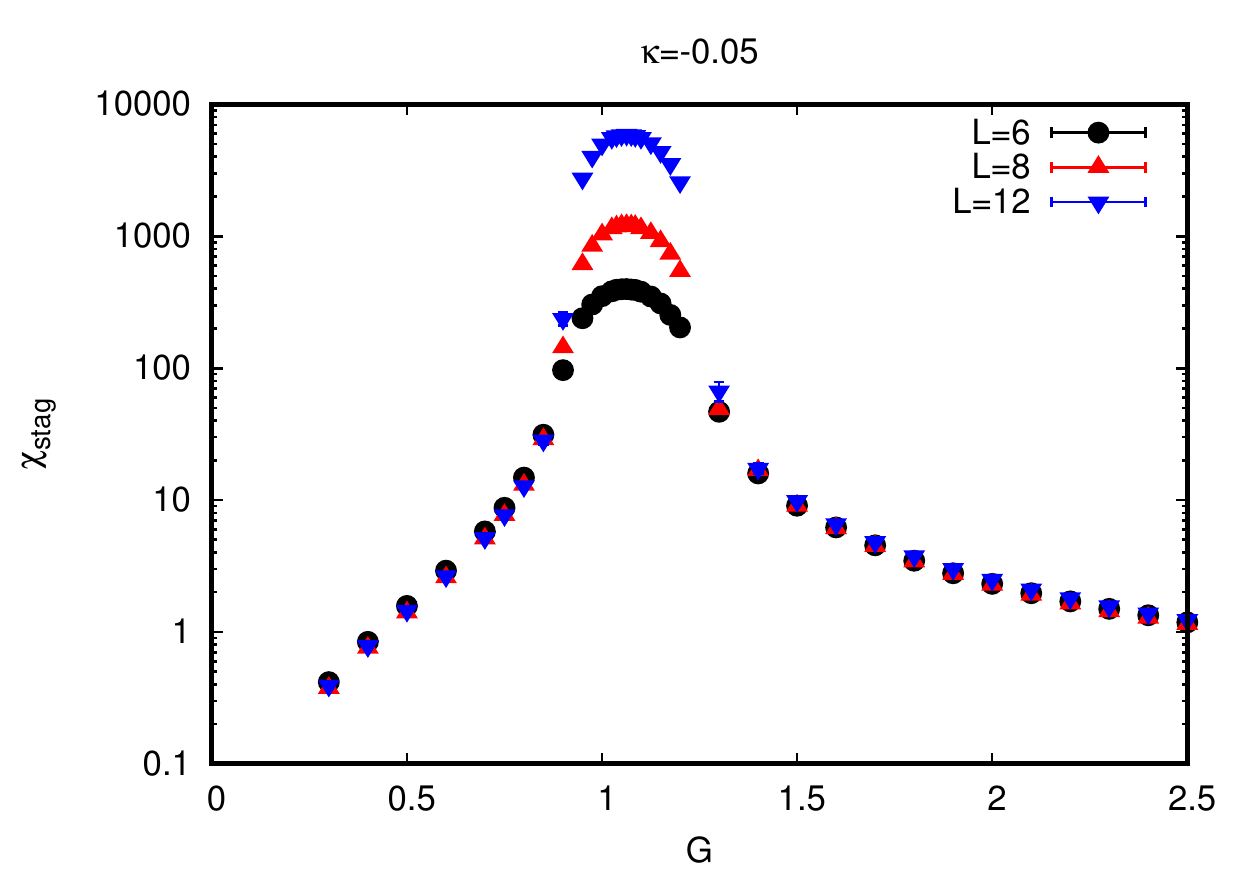}
  \caption{$\chi_{\text{stag}}$ vs $G$ at $\kappa=-0.05$ for $L=6$, 8 and 12.}
  \label{fig:sus_-0.05}
\end{figure}

Since $\kappa<0$ generates additional antiferromagnetic terms in the effective fermion action we expect this
bilinear phase to survive in the $\kappa<0$ region of the phase diagram. This is confirmed in our calculations.
Figure~\ref{fig:sus_-0.05} shows a similar susceptibility plot for $\kappa=-0.05$, in which the width of the broken phase
\textit{increases} while the peak height continues to scale linearly with the volume indicating the presence
of an antiferromagetic bilinear condensate.

\begin{figure}[tbp]
  \centering
  \includegraphics[width=0.48\textwidth]{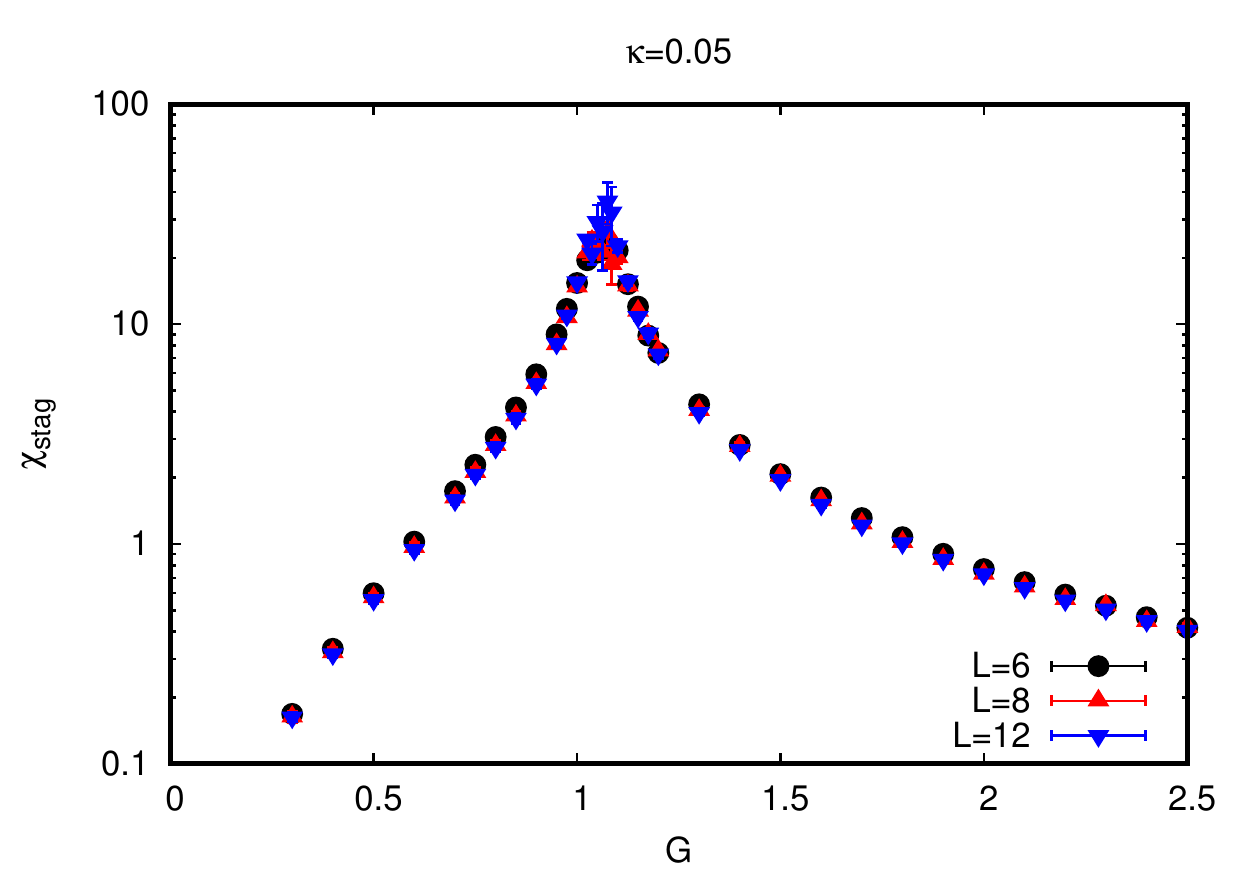}
  \caption{$\chi_{\text{stag}}$ vs $G$ at $\kappa=0.05$ for $L=6$, 8 and 12.}
  \label{fig:sus_0.05}
\end{figure}

\begin{figure}
  \centering
  \includegraphics[width=0.48\textwidth]{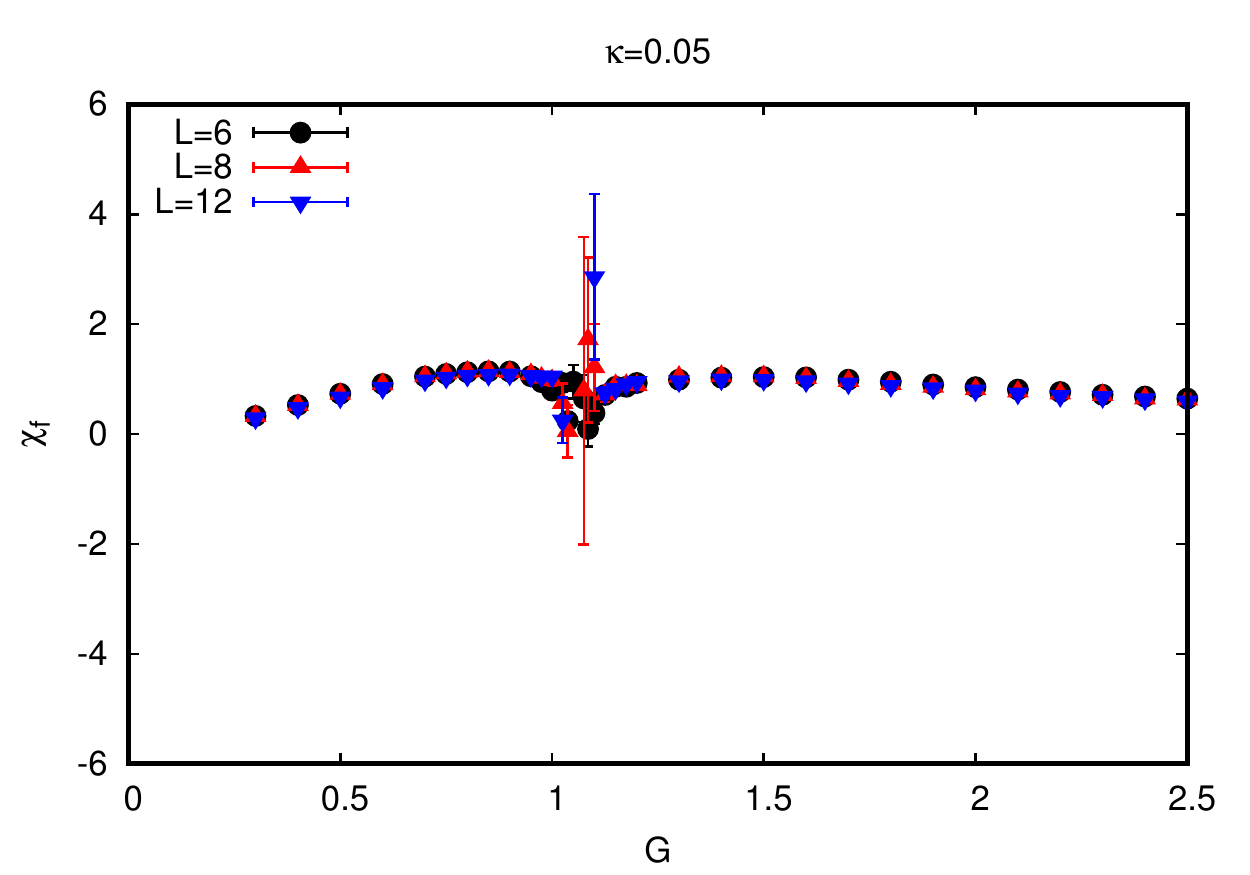}
  \caption{The ferromagnetic susceptibility $\chi_{\text{f}}$ vs $G$ at $\kappa=0.05$ for $L=6$, 8 and 12.  Unlike the other susceptibility plots, the y-axis scale is not logarithmic.}
  \label{fig:sus_s_0.005}
\end{figure}

The situation changes for $\kappa>0$. Fig.~\ref{fig:sus_0.05} shows the susceptibility $\chi_{\text{stag}}$ for
$\kappa=0.05$. While a peak is still observed for essentially the same value of $G$ the height of this
peak no longer scales with the volume.
Since $\kappa>0$ induces ferromagnetic terms in the action we also examine the
associated ferromagnetic susceptibility
\begin{equation}
  \chi_{\text{f}} = \frac{1}{V} \sum_{x,y,a,b} \vev{\psi^a(x)\psi^b(x) \psi^a(y)\psi^b(y)}.
\end{equation}
This is plotted in Fig.~\ref{fig:sus_s_0.005} for $\kappa=0.05$, which shows no evidence of ferromagnetic ordering at this value of $\kappa$.
In the appendix we show that $\kappa = 0.1$ is sufficiently large to produce a ferromagnetic phase.

\begin{figure}
  \centering
  \includegraphics[width=0.48\textwidth]{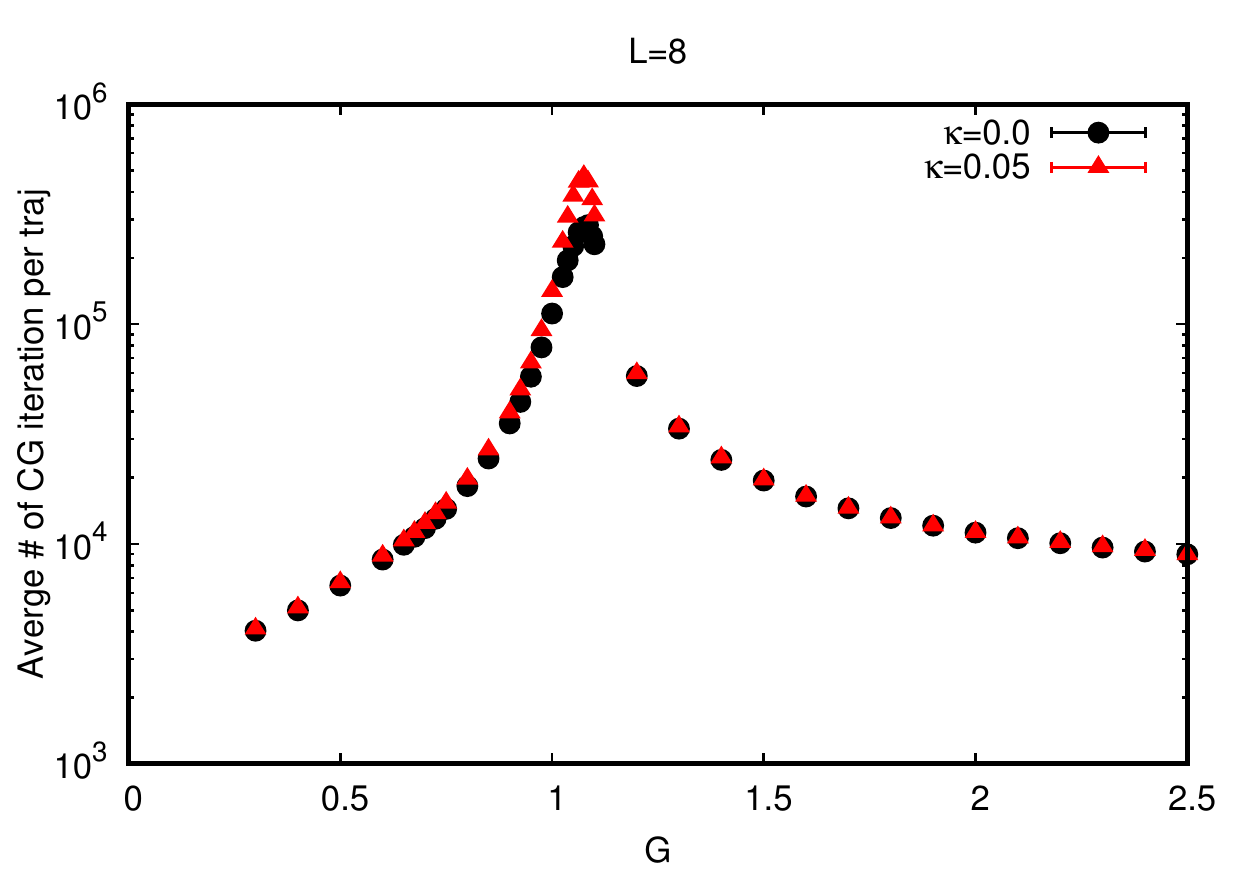}
  \caption{Average number of CG iterations for Dirac operator inversions on $L = 8$ lattices, plotted vs $G$ for $\kappa = 0$ and 0.05.}\label{fig:CG}
\end{figure}

The lack of scaling of the $\chi_{\text{stag}}$ peak with volume at $\kappa=0.05$ might suggest that the
system is no longer critical at this point. This is not the case. Figure~\ref{fig:CG} shows the number of
conjugate gradient (CG) iterations needed for Dirac operator inversions at $\kappa=0$ and $\kappa=0.05$
as a function of $G$ for $L=8$. This quantity is a proxy for the fermion correlation length in the system.
The peak at $\kappa=0.05$ is significantly greater than at $\kappa=0$. Furthermore we have observed
that it increases strongly with lattice size rendering it very difficult to run computations for $L \geq 16$. Our conclusion
is that there is still a phase transition around $G\approx 1.05$ for small positive $\kappa$ but no sign of
a bilinear condensate. We will reinforce this conclusion in the next section where we will perform an analysis
of bilinear vevs versus external symmetry breaking sources.

\begin{figure}[tbp]
  \centering
  \includegraphics[width=0.48\textwidth]{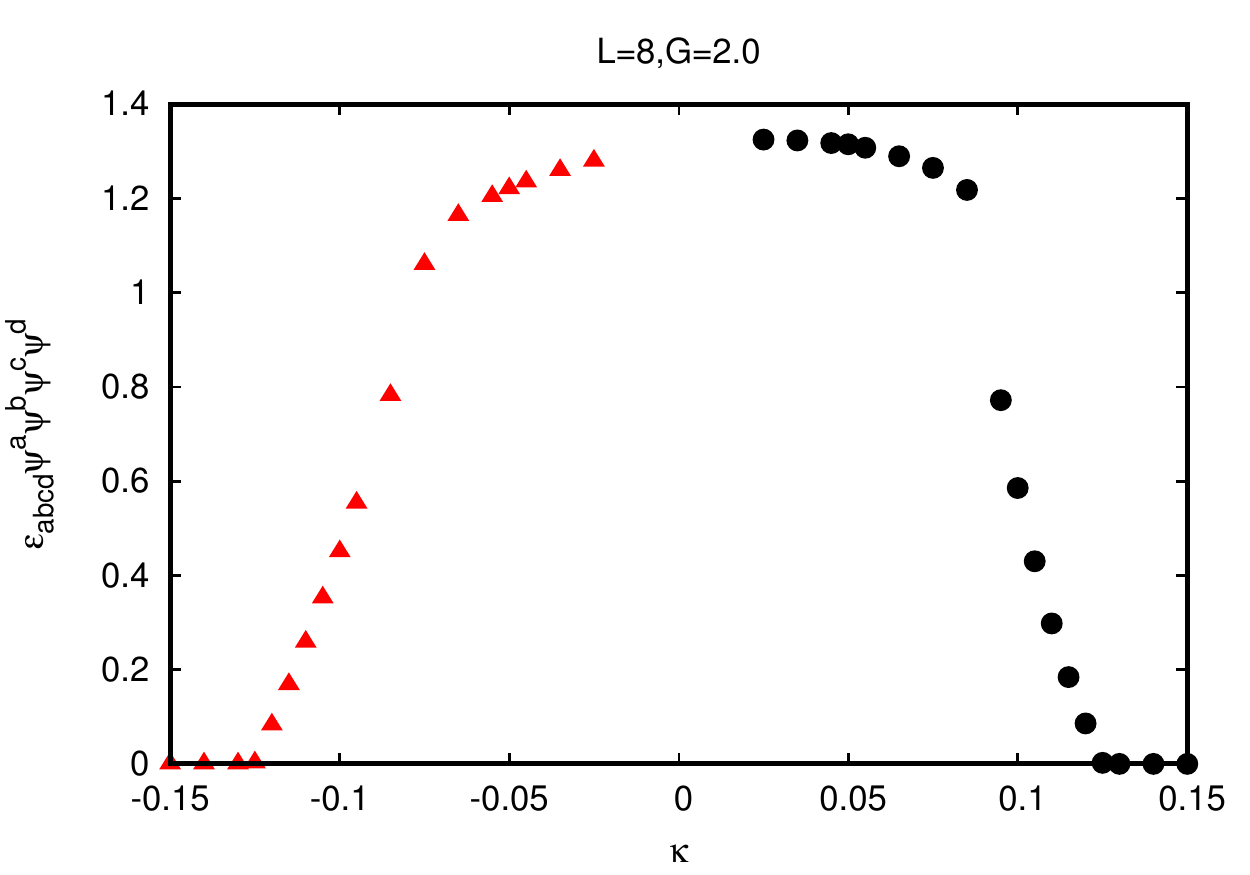}
  \caption{Four fermion condensate at $G=2$ vs $\kappa$ for $L=8$.}
  \label{fig:fourvsk}
\end{figure}

\begin{figure}
  \centering
  \includegraphics[width=0.48\textwidth]{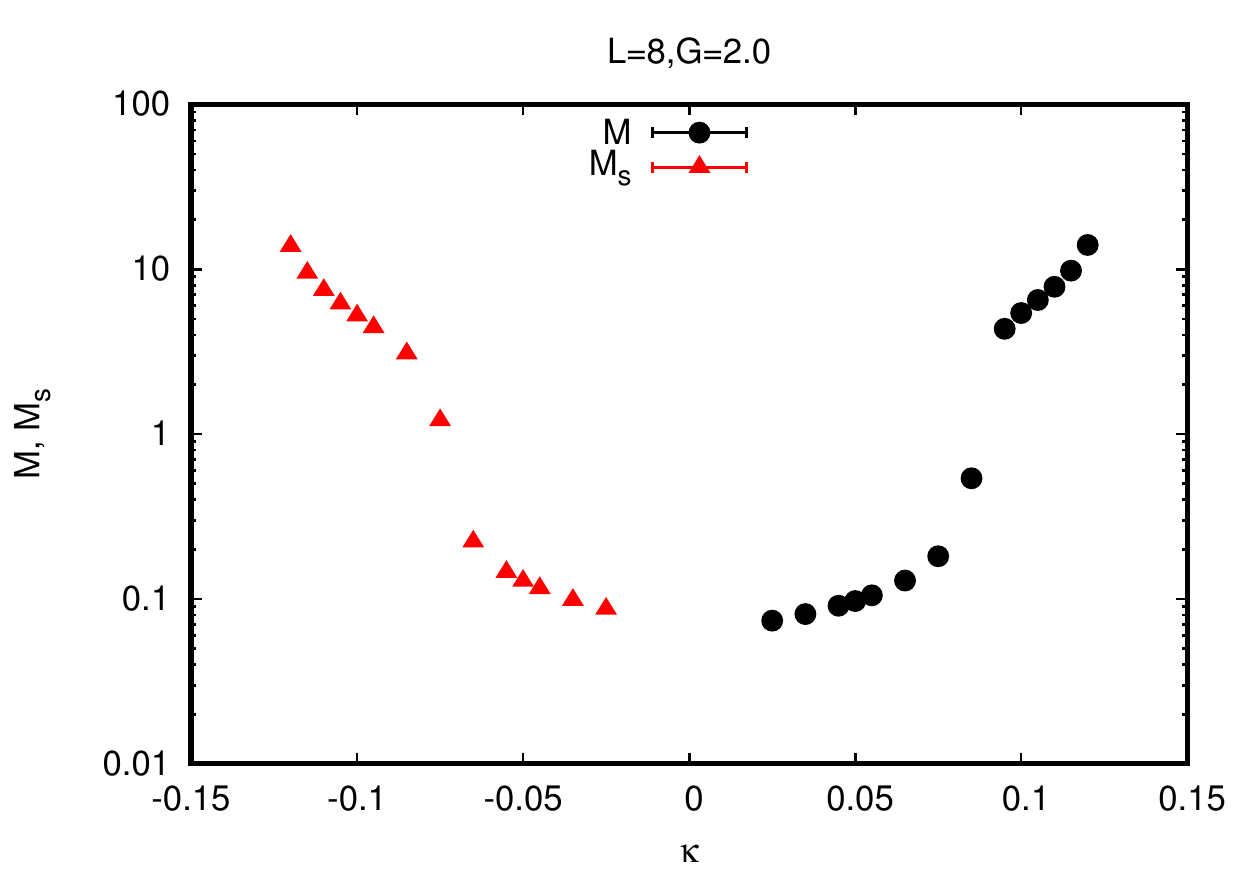}
  \caption{Ferromagnetic and staggered magnetizations at $G=2$ vs $\kappa$ for $L=8$.}
  \label{fig:mag_s}
\end{figure}

It is interesting to investigate the phase diagram away from the critical region. Figure~\ref{fig:fourvsk} shows the four fermion
condensate vs $\kappa$ at $G = 2$, which vanishes at $|\kappa|=\frac{1}{8}$ as
expected by stability arguments. The structure of the curve suggests that there may be a phase transition
at $\kappa\approx 0.085$ from a four fermion condensate to a ferromagnetic condensate. This is illustrated by Fig.~\ref{fig:mag_s} where for $\kappa>0$ we
show the magnetization
\begin{equation}
  \label{eq:mag}
  M = \frac{1}{V} \vev{\left| \sum_x \sum_{a < b} \sigma^{+}_{ab}(x) \right|}.
\end{equation}
The behavior near $\kappa\approx -0.085$ in Fig.~\ref{fig:mag_s}
shows a similar transition from four fermion condensate to antiferromagnetic phase.  For $\kappa < 0$ we add the usual parity factor $\epsilon(x)$ to define the staggered magnetization $M_s$.

\section{\label{sec:bilin}Fermion Bilinears}
\begin{figure}[tbp]
\centering
  \includegraphics[width=0.48\textwidth]{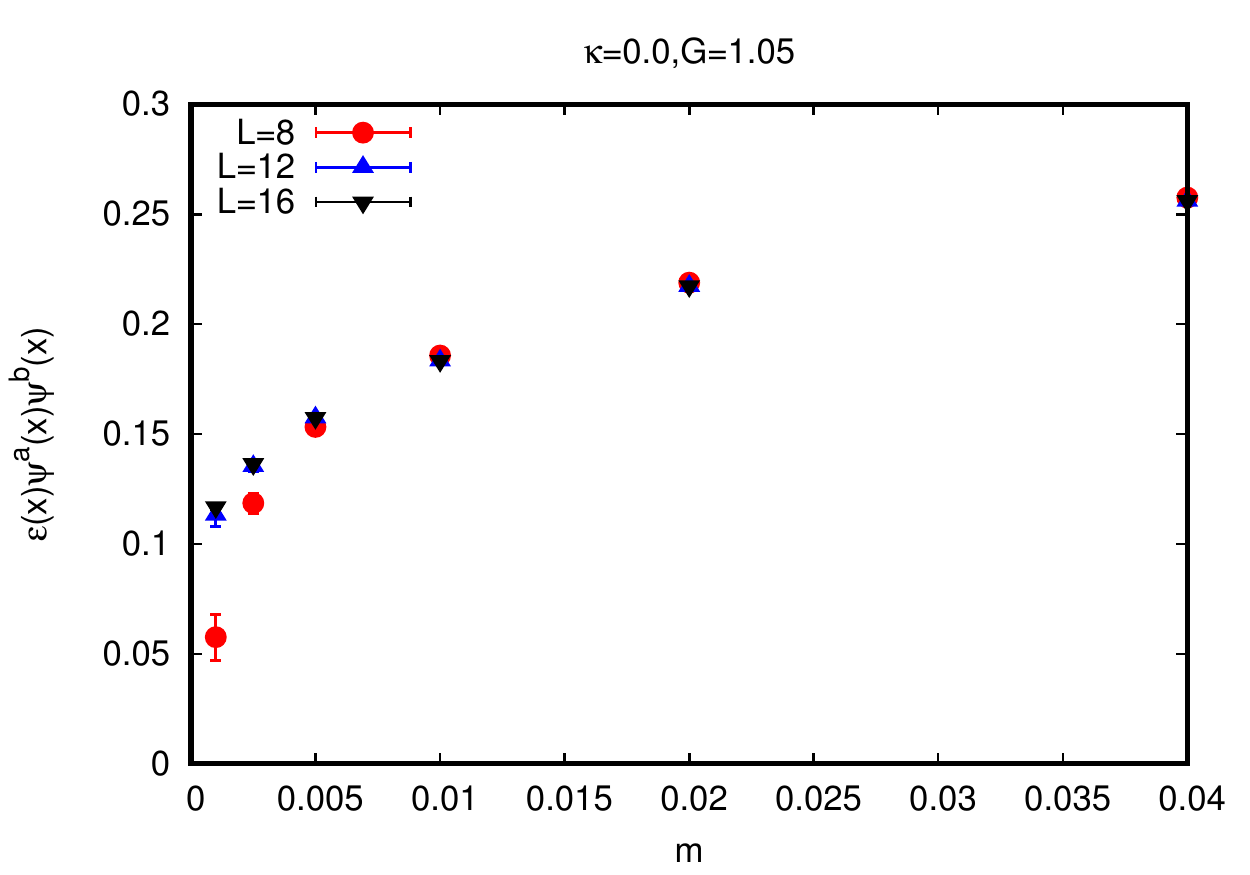}
  \caption{Antiferromagnetic bilinear condensate vs $m_2$ (with $m_1 = 0$) at $\left(\kappa, G \right) = (0, 1.05)$ for $L = 8$, 12 and 16.}
  \label{fig:bilin}
\end{figure}

In this section we add source terms to the action that explicitly break both the $SO(4)$ and $Z_2$ symmetries
and, by examining the volume dependence of various bilinear vevs as the sources are sent to zero, address the question of whether spontaneous symmetry breaking occurs in the system.  The source terms take the form
\begin{equation}
  \label{eq:d}
  \delta S = \sum_{x,a,b} (m_1 + m_2 \epsilon(x) ) [ \psi^a(x)\psi^{b}(x) ] \Sigma^{ab}_{+}
\end{equation}
where the $SO(4)$ symmetry breaking source $\Sigma^{ab}_{+}$ is
\begin{equation}
  \Sigma^{ab}_{+} = \left(\begin{array}{cc}i\sigma_2 & 0\\
                                           0         & i\sigma_2\end{array}\right).
\end{equation}
For $\kappa =0$ we find evidence in favor of antiferromagnetic ordering consistent with the volume scaling of the susceptibility $\chi_{\text{stag}}$.
The antiferromagnetic bilinear vev $\vev{\epsilon(x) \psi^a(x) \psi^b(x)}$ plotted in Fig.~\ref{fig:bilin} (with $m_1=0$) picks up a non-zero value in the limit $m_2 \to 0$, $L \to \infty$ signaling spontaneous symmetry breaking. The data correspond
to runs at the peak in the susceptibility $G=1.05$, and similar results are found throughout the region
$0.95 \lesssim G \lesssim 1.15$. This confirms the presence of the condensate inferred from the linear volume scaling of the
susceptibility reported in the previous section.

\begin{figure}[tbp]
  \centering
  \includegraphics[width=0.48\textwidth]{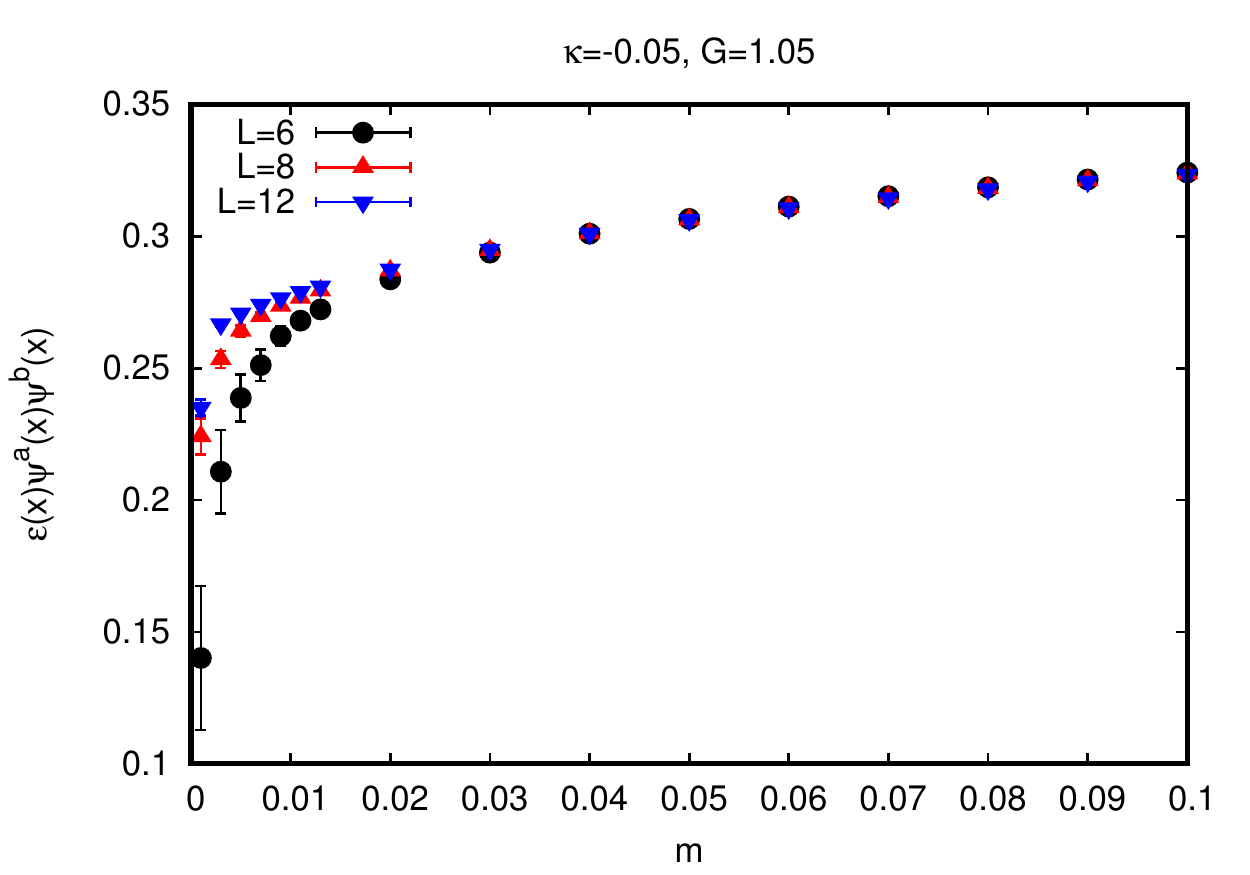}
  \caption{Antiferromagnetic bilinear condensate vs $m_2 = m_1$ at $\left(\kappa, G \right) = (-0.05, 1.05)$ for $L = 6$, 8 and 12.}
  \label{fig:af_bi}
\end{figure}

For $\kappa<0$ the picture is similar with Fig.~\ref{fig:af_bi} showing the same vev vs $m_2=m_1$ for $\kappa = -0.05$ at the same $G = 1.05$.
(Recall from Figs.~\ref{fig:sus_0.0}--\ref{fig:sus_0.05} that the center of the peak in $\chi_{\text{stag}}$ moves only very slowly for $|\kappa| \leq 0.05$.)
The increase in vev with larger volumes at small $m_2$ is again very consistent with the presence of a non-zero condensate in the thermodynamic limit.
The magnitude of this condensate at $\kappa = -0.05$ is clearly larger than at $\kappa=0$.

\begin{figure*}[tbp]
  \centering
  \includegraphics[width=0.48\textwidth]{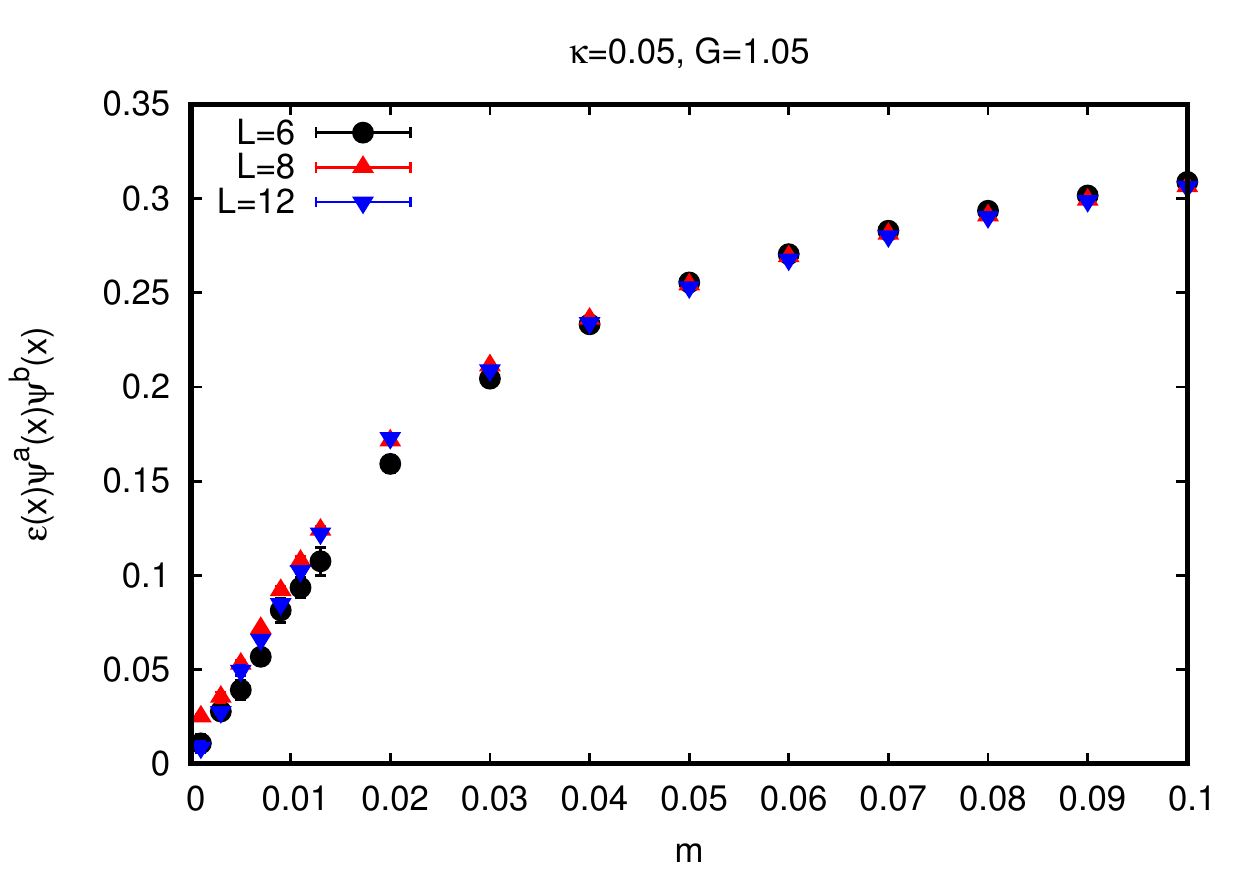} \hfill \includegraphics[width=0.48\textwidth]{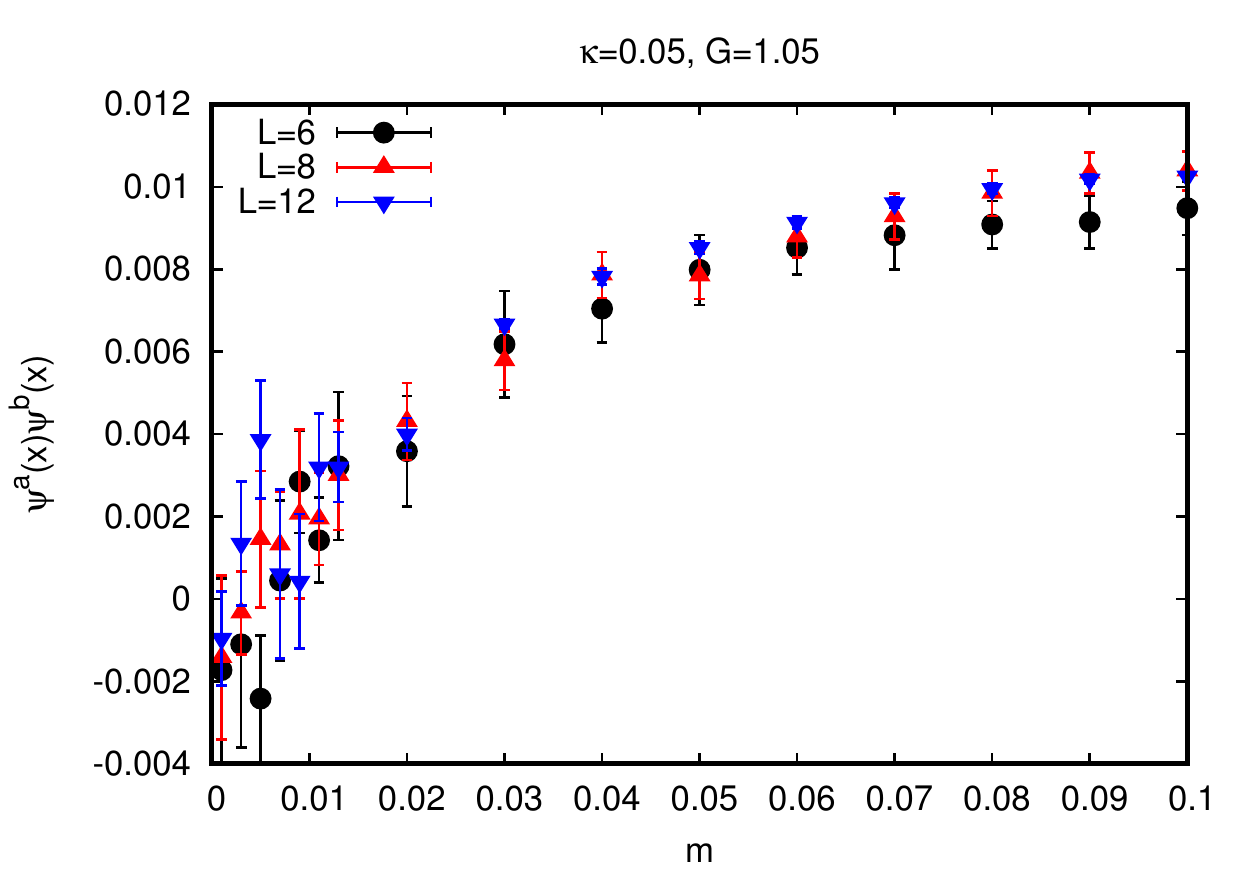}
  \caption{Antiferromagnetic (left) and ferromagnetic (right) bilinear condensates vs $m_2 = m_1$ at $\left(\kappa, G \right) = (0.05, 1.05)$ for $L=6$, 8 and 12.}
  \label{fig:cond_0.05}
\end{figure*}

The situation for $\kappa>0$ is quite different. Figure~\ref{fig:cond_0.05} shows plots of both antiferromagnetic
and ferromagnetic bilinear vevs at $\left(\kappa, G \right) = (0.05, 1.05)$ for several lattice volumes.
These plots show no sign of a condensate as the source terms are removed in the thermodynamic limit.
Broken phases thus seem to be evaded for small $\kappa>0$.
In the appendix we include results for larger $\kappa \geq 0.085$.
While we observe a similar absence of bilinear condensates at $\kappa = 0.085$, the expected ferromagnetic phase does clearly appear for $\kappa \approx 0.1$ and we are able to set loose bounds on the range $\kappa_1 < \kappa < \kappa_2$ within which there appears to be a direct PMW--PMS transition, namely $0 < \kappa_1 < 0.05$ while $0.085 < \kappa_2 < 0.125$.

\section{\label{sec:diagram}Resulting phase diagram}
\begin{figure}[tbp]
  \centering
  \includegraphics[width=0.48\textwidth]{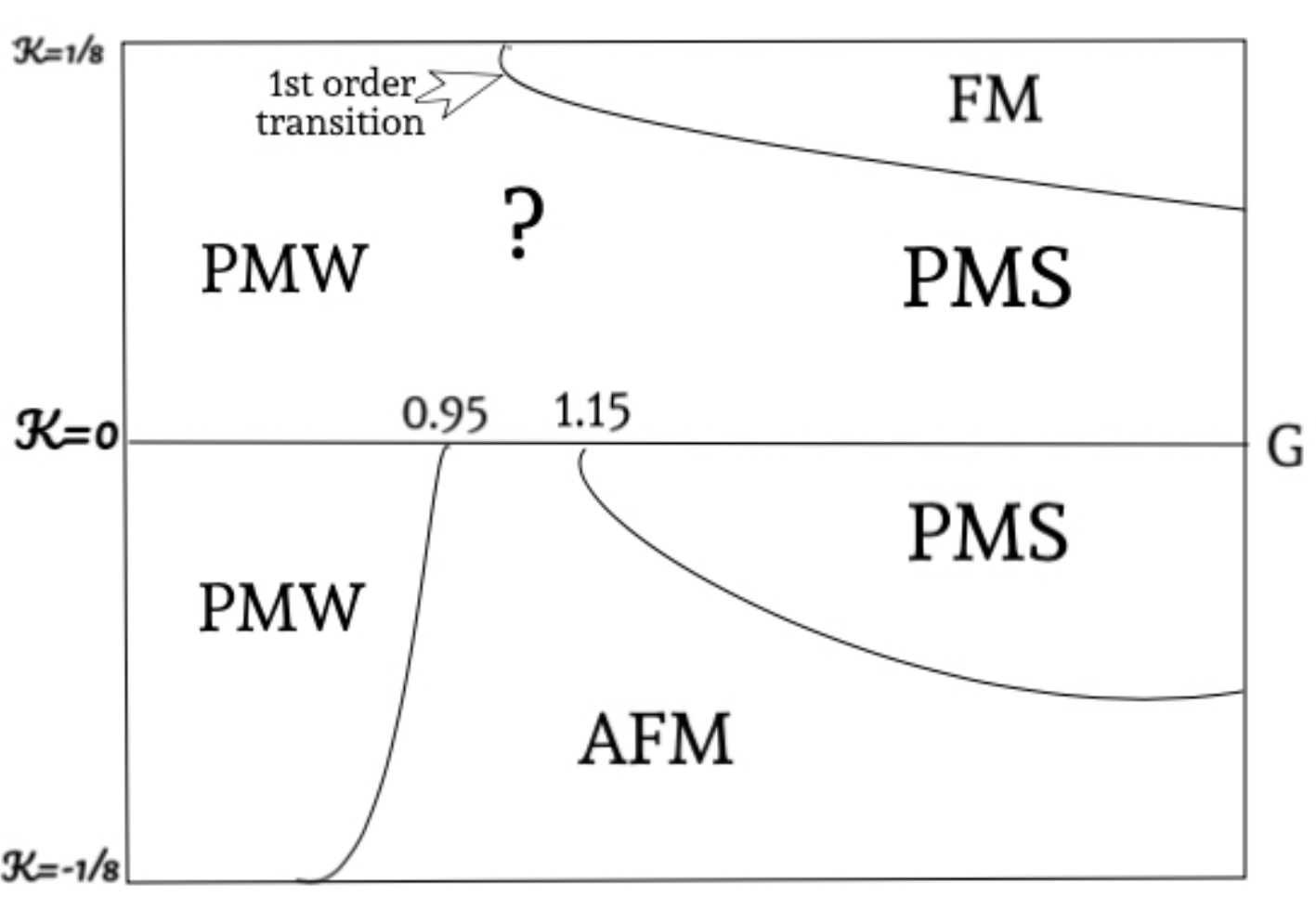}
  \caption{Sketch of the phase diagram in the $\left(\kappa, G \right)$ plane.}
  \label{fig:phasediagram}
\end{figure}

Putting this all together we sketch the
phase diagram in Fig.~\ref{fig:phasediagram}. For small $G$ the system is disordered and the fermions
massless. For large $G$ and small $\kappa$ we see a four
fermion condensate as before. As $\kappa$ increases in magnitude one expects a transition to either
a ferromagnetic ($\kappa>0$) or antiferromagnetic ($\kappa<0$) phase for sufficiently large $G$.
However, for small positive $\kappa$ close to $G=1.05$, while
we observe no sign of a bilinear condensate there are strong indications of critical slowing down and
a large fermion correlation length. Since the weak and strong coupling phases \textit{cannot} be analytically
connected (one is massless while in the other the fermions acquire a mass) there must be at least
one phase
transition between them. Unlike the situation for $\kappa\le 0$ we see no evidence for an intermediate broken-symmetry
phase in this region and hence the simplest conclusion is that a single phase transition separates the two
symmetric phases. Thus far we have seen no sign of first order behavior so this transition appears to
be continuous.

\section{\label{sec:conc}Summary and Conclusions}
In this paper we have reported on investigations of the phase diagram of a four-dimensional lattice Higgs-Yukawa model comprising four reduced staggered fermions interacting with a scalar field transforming in the self-dual representation of a global $SO(4)$ symmetry.
This extends recent work on a related four fermion model in which a massless symmetric phase is separated from a massive symmetric phase by a narrow broken symmetry phase characterized by a small antiferromagnetic bilinear fermion condensate~\cite{Ayyar:2016lxq, Ayyar:2016nqh, Catterall:2016dzf, Schaich:2017czc}.

Our main result is evidence that this broken phase may be eliminated in the generalized phase diagram by tuning the hopping parameter in the scalar kinetic term.
This should not be too surprising since the ferromagnetic ordering favored by $\kappa > 0$ counteracts the antiferromagnetic ordering observed for $\kappa \leq 0$.
There is then a range of positive $\kappa_1 < \kappa < \kappa_2$ throughout which the massless and massive symmetric phases appear to be separated by a single phase transition.
Since no order parameter distinguishes the two phases this transition is not of a conventional Landau-Ginzburg type.
Ref.~\cite{Catterall:2017ogi} argues in a related continuum model that the transition may be driven instead by topological defects.
It would be fascinating to investigate whether these topological defects could be seen in numerical calculations.

Future work will also focus on better constraining the values of $\kappa_1$ and $\kappa_2$ between which we observe the direct PMW--PMS transition.
Our current results suffice to establish that these two points are well separated, $0 < \kappa_1 < 0.05$ while $0.085 < \kappa_2 < 0.125$, but neither is very precisely determined yet.
It is also important to measure more observables in order to search for non-trivial scaling behavior associated with this transition.
The lack of scaling that we observe for the susceptibility $\chi_{\text{stag}}$ at the phase boundary in Fig.~\ref{fig:sus_0.05} currently suggests that the scaling dimension of the bilinear fermion operator would be greater than two at any putative new critical point.

Clearly the possibility of realizing new fixed points in strongly interacting fermionic systems in four dimensions
is of great interest and we hope our results stimulate further work in this area.

\acknowledgments
We thank Shailesh Chandrasekharan, Jarno Rantaharju and Anna Hasenfratz for pleasant and productive conversations.
This work is supported in part by the U.S.\ Department of Energy (DOE), Office of Science, Office of High Energy Physics, under Award Number {DE-SC0009998}.  Numerical computations were performed at Fermilab using USQCD resources funded by the DOE Office of Science.

\newpage 

\section*{Appendix}
\begin{figure*}[tbp] 
  \centering
  \includegraphics[width=0.48\textwidth]{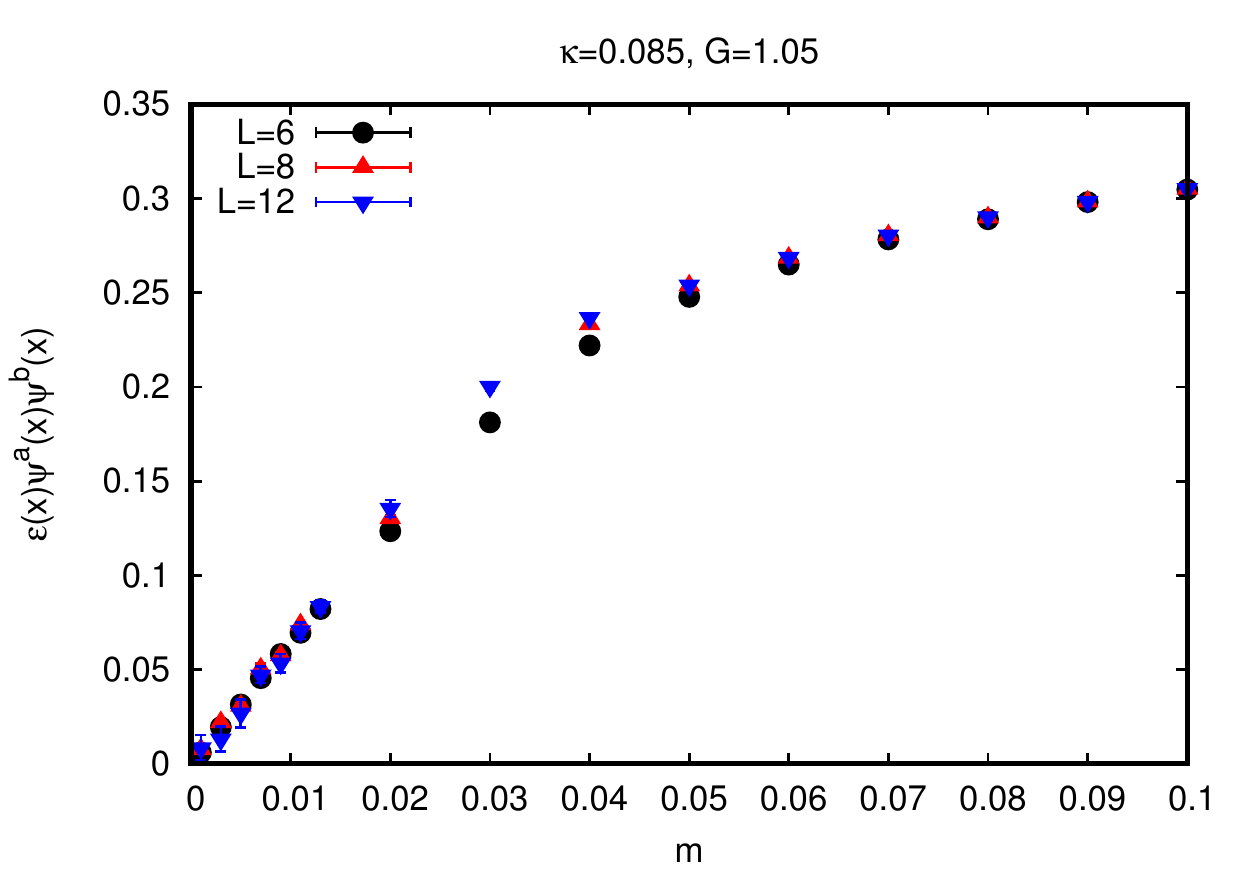} \hfill \includegraphics[width=0.48\textwidth]{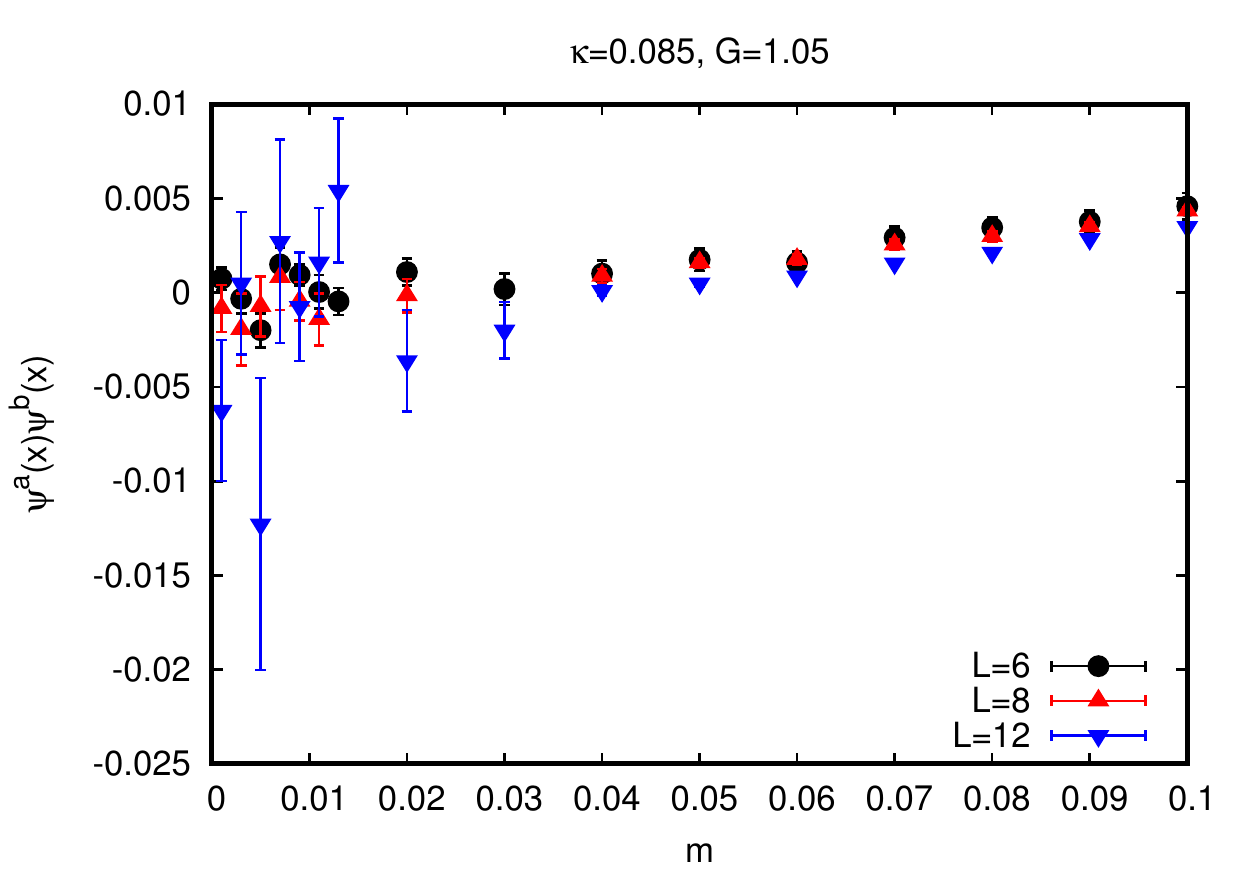}
  \caption{Antiferromagnetic (left) and ferromagnetic (right) bilinear condensates vs $m_2 = m_1$ at $\left(\kappa, G \right) = (0.085, 1.05)$ for $L=6$, 8 and 12.}
  \label{fig:cond_0.085}
\end{figure*}

In this appendix we collect some additional results for larger $\kappa > 0.05$, both to strengthen our conclusions that there is no bilinear phase for a range of positive $\kappa$ and to confirm that a ferromagnetic phase does appear once $\kappa$ and $G$ are sufficiently large.
First, in Fig.~\ref{fig:cond_0.085} we consider $\kappa = 0.085$, around the potential transition identified in Figs.~\ref{fig:fourvsk} and \ref{fig:mag_s}.
Whereas those earlier figures considered $G = 2$, here we use the same $G = 1.05$ as Fig.~\ref{fig:cond_0.05} for $\kappa = 0.05$.
We again observe an absence of spontaneous symmetry breaking, with the antiferromagnetic and ferromagnetic bilinear condensates both vanishing as the symmetry-breaking source terms are removed, with no visible dependence on the lattice volume.

\begin{figure}[tbp]
  \centering
  \includegraphics[width=0.48\textwidth]{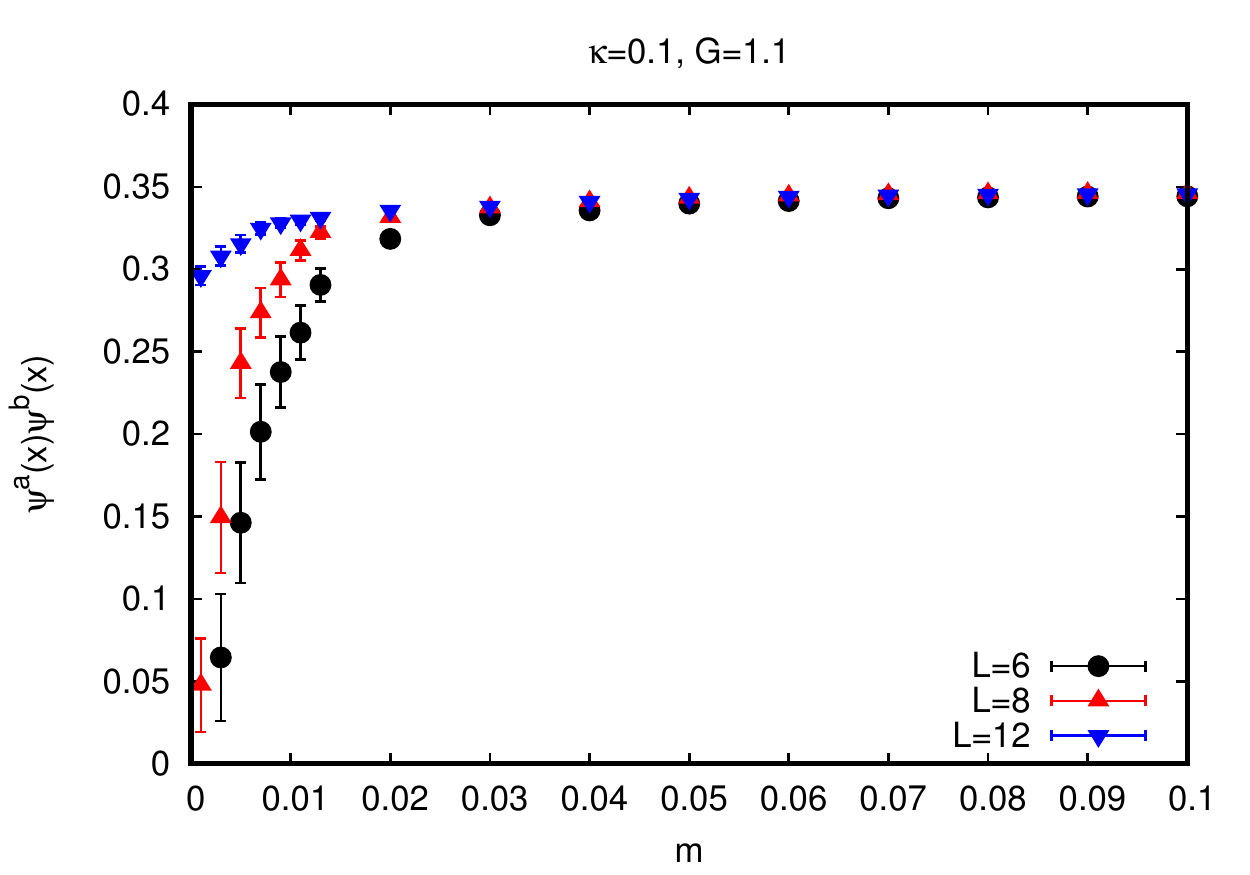}
  \caption{Ferromagnetic bilinear condensate vs $m_2=m_1$ at $\left(\kappa, G \right) = (0.1, 1.1)$ for $L=6$, 8 and 12.}
  \label{fig:cond_0.1}
\end{figure}

\begin{figure}[tbp]
  \centering
  \includegraphics[width=0.48\textwidth]{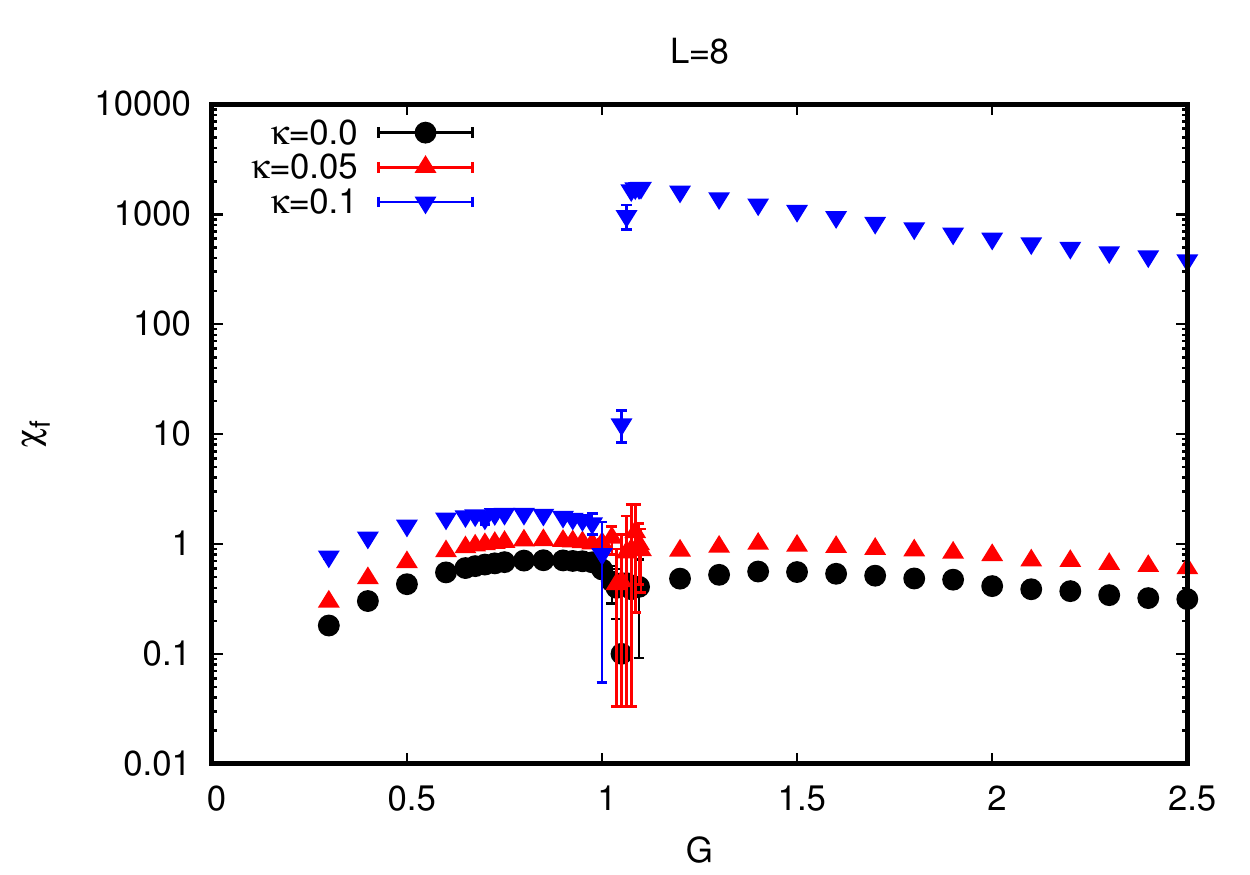}
  \caption{The $L = 8$ ferromagnetic susceptibility $\chi_{\text{f}}$ vs $G$ for $\kappa = 0$, 0.05 and 0.1.}
  \label{fig:sus_0.1}
\end{figure}

\begin{figure}[tbp]
  \centering
  \includegraphics[width=0.48\textwidth]{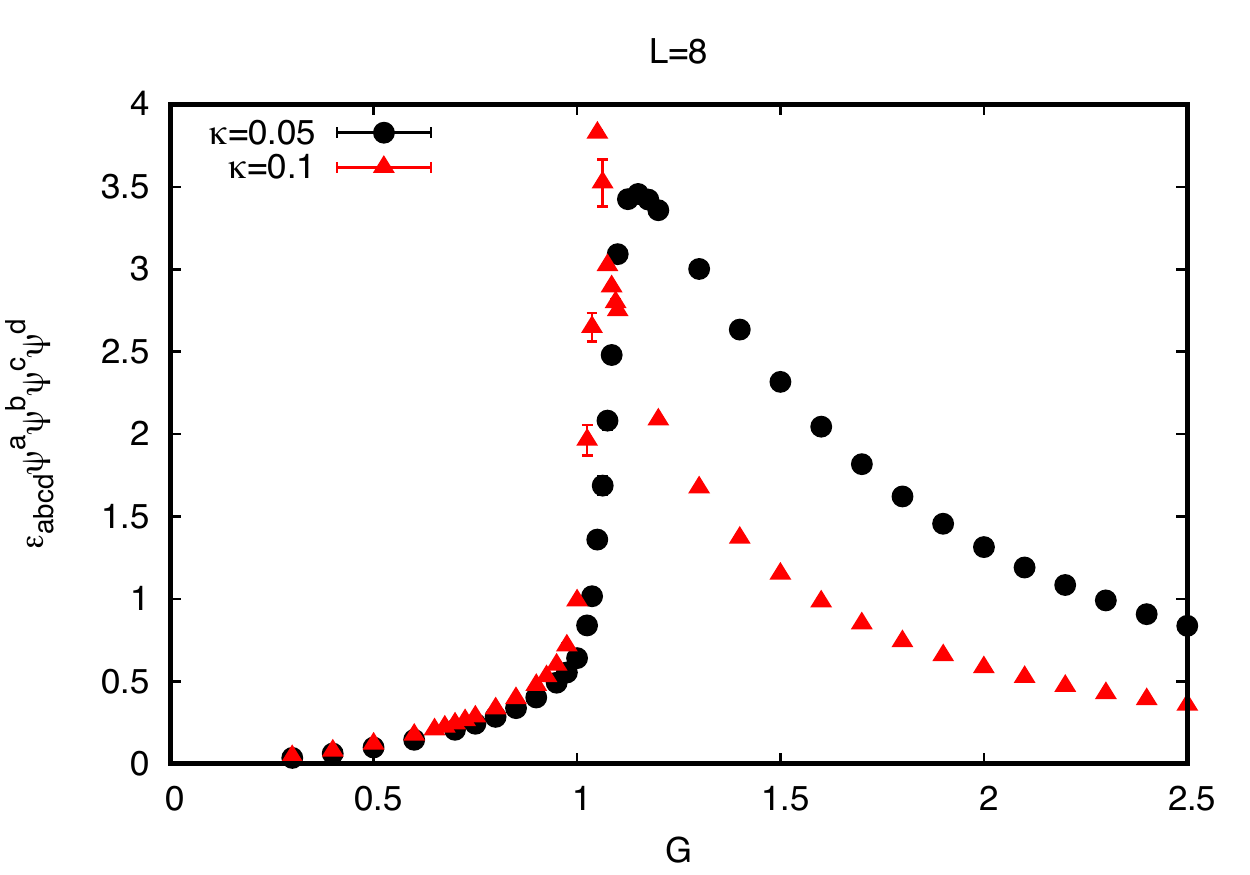}
  \caption{Four fermion condensate on vs $G$ for $L = 8$, comparing $\kappa = 0.5$ and 0.1.}
  \label{fig:fourvsG}
\end{figure}

The situation is qualitatively different in Fig.~\ref{fig:cond_0.1}, which considers $\kappa = 0.1$ (at $G = 1.1$) and shows clear signs of a non-zero ferromagnetic condensate in the $L \to \infty$ limit.
In Fig.~\ref{fig:sus_0.1} we compare the ferromagnetic susceptibility $\chi_{\text{f}}$ for three different $\kappa = 0$, 0.05 and 0.1.
While this susceptibility is uniformly small for $\kappa = 0$ and 0.05, the larger $\kappa = 0.1$ produces a strong jump to a large value for $G \gtrsim 1.1$, suggesting a first-order transition into the ferromagnetic phase.

Finally, Fig.~\ref{fig:fourvsG} compares the four-fermion condensate vs $G$ for $\kappa = 0.05$ and 0.1.
Although the larger value of $\kappa$ significantly reduces the four-fermion condensate for large $G \gtrsim 1.2$ (as previously shown in Fig.~\ref{fig:fourvsk}), there is a very narrow peak around $G \approx 1.05$.
This may suggest that the system still transitions directly from the PMW phase into the PMS phase before undergoing a second transition into the ferromagnetic phase.
We are therefore not yet able to set tighter constraints than $0.085 < \kappa_2 < 0.125$ on the upper boundary of the direct PMW--PMS transition.
This region of the phase diagram appears rather complicated, though Fig.~\ref{fig:sus_0.1} makes it clear that the ferromagnetic phase persists to large $G$ rather than being a narrow intermediate phase of the sort we see for $\kappa \leq 0$.
This is reflected in our sketch of the phase diagram, Fig.~\ref{fig:phasediagram}.

\newpage 

\raggedright
\bibliography{mybib}

\begin{thebibliography}{21}%
\makeatletter
\providecommand \@ifxundefined [1]{%
 \@ifx{#1\undefined}
}%
\providecommand \@ifnum [1]{%
 \ifnum #1\expandafter \@firstoftwo
 \else \expandafter \@secondoftwo
 \fi
}%
\providecommand \@ifx [1]{%
 \ifx #1\expandafter \@firstoftwo
 \else \expandafter \@secondoftwo
 \fi
}%
\providecommand \natexlab [1]{#1}%
\providecommand \enquote  [1]{``#1''}%
\providecommand \bibnamefont  [1]{#1}%
\providecommand \bibfnamefont [1]{#1}%
\providecommand \citenamefont [1]{#1}%
\providecommand \href@noop [0]{\@secondoftwo}%
\providecommand \href [0]{\begingroup \@sanitize@url \@href}%
\providecommand \@href[1]{\@@startlink{#1}\@@href}%
\providecommand \@@href[1]{\endgroup#1\@@endlink}%
\providecommand \@sanitize@url [0]{\catcode `\\12\catcode `\$12\catcode
  `\&12\catcode `\#12\catcode `\^12\catcode `\_12\catcode `\%12\relax}%
\providecommand \@@startlink[1]{}%
\providecommand \@@endlink[0]{}%
\providecommand \url  [0]{\begingroup\@sanitize@url \@url }%
\providecommand \@url [1]{\endgroup\@href {#1}{\urlprefix }}%
\providecommand \urlprefix  [0]{URL }%
\providecommand \Eprint [0]{\href }%
\providecommand \doibase [0]{http://dx.doi.org/}%
\providecommand \selectlanguage [0]{\@gobble}%
\providecommand \bibinfo  [0]{\@secondoftwo}%
\providecommand \bibfield  [0]{\@secondoftwo}%
\providecommand \translation [1]{[#1]}%
\providecommand \BibitemOpen [0]{}%
\providecommand \bibitemStop [0]{}%
\providecommand \bibitemNoStop [0]{.\EOS\space}%
\providecommand \EOS [0]{\spacefactor3000\relax}%
\providecommand \BibitemShut  [1]{\csname bibitem#1\endcsname}%
\let\auto@bib@innerbib\@empty
\bibitem [{\citenamefont {Ayyar}\ and\ \citenamefont
  {Chandrasekharan}(2015)}]{Ayyar:2014eua}%
  \BibitemOpen
  \bibfield  {author} {\bibinfo {author} {\bibfnamefont {V.}~\bibnamefont
  {Ayyar}}\ and\ \bibinfo {author} {\bibfnamefont {S.}~\bibnamefont
  {Chandrasekharan}},\ }\bibfield  {title} {\enquote {\bibinfo {title}
  {{Massive fermions without fermion bilinear condensates}},}\ }\href {\doibase
  10.1103/PhysRevD.91.065035} {\bibfield  {journal} {\bibinfo  {journal} {Phys.
  Rev.}\ }\textbf {\bibinfo {volume} {D91}},\ \bibinfo {pages} {065035}
  (\bibinfo {year} {2015})},\ \Eprint {http://arxiv.org/abs/1410.6474}
  {arXiv:1410.6474} \BibitemShut {NoStop}%
\bibitem [{\citenamefont {Ayyar}\ and\ \citenamefont
  {Chandrasekharan}(2016{\natexlab{a}})}]{Ayyar:2015lrd}%
  \BibitemOpen
  \bibfield  {author} {\bibinfo {author} {\bibfnamefont {V.}~\bibnamefont
  {Ayyar}}\ and\ \bibinfo {author} {\bibfnamefont {S.}~\bibnamefont
  {Chandrasekharan}},\ }\bibfield  {title} {\enquote {\bibinfo {title} {{Origin
  of fermion masses without spontaneous symmetry breaking}},}\ }\href {\doibase
  10.1103/PhysRevD.93.081701} {\bibfield  {journal} {\bibinfo  {journal} {Phys.
  Rev.}\ }\textbf {\bibinfo {volume} {D93}},\ \bibinfo {pages} {081701}
  (\bibinfo {year} {2016}{\natexlab{a}})},\ \Eprint
  {http://arxiv.org/abs/1511.09071} {arXiv:1511.09071} \BibitemShut {NoStop}%
\bibitem [{\citenamefont {Catterall}(2016)}]{Catterall:2015zua}%
  \BibitemOpen
  \bibfield  {author} {\bibinfo {author} {\bibfnamefont {S.}~\bibnamefont
  {Catterall}},\ }\bibfield  {title} {\enquote {\bibinfo {title} {{Fermion mass
  without symmetry breaking}},}\ }\href {\doibase 10.1007/JHEP01(2016)121}
  {\bibfield  {journal} {\bibinfo  {journal} {JHEP}\ }\textbf {\bibinfo
  {volume} {01}},\ \bibinfo {pages} {121} (\bibinfo {year} {2016})},\ \Eprint
  {http://arxiv.org/abs/1510.04153} {arXiv:1510.04153} \BibitemShut {NoStop}%
\bibitem [{\citenamefont {He}\ \emph {et~al.}(2016)\citenamefont {He},
  \citenamefont {Wu}, \citenamefont {You}, \citenamefont {Xu}, \citenamefont
  {Meng},\ and\ \citenamefont {Lu}}]{He:2016sbs}%
  \BibitemOpen
  \bibfield  {author} {\bibinfo {author} {\bibfnamefont {Y.-Y.}\ \bibnamefont
  {He}}, \bibinfo {author} {\bibfnamefont {H.-Q.}\ \bibnamefont {Wu}}, \bibinfo
  {author} {\bibfnamefont {Y.-Z.}\ \bibnamefont {You}}, \bibinfo {author}
  {\bibfnamefont {C.}~\bibnamefont {Xu}}, \bibinfo {author} {\bibfnamefont
  {Z.~Y.}\ \bibnamefont {Meng}}, \ and\ \bibinfo {author} {\bibfnamefont
  {Z.-Y.}\ \bibnamefont {Lu}},\ }\bibfield  {title} {\enquote {\bibinfo {title}
  {{Quantum critical point of Dirac fermion mass generation without spontaneous
  symmetry breaking}},}\ }\href {\doibase 10.1103/PhysRevB.94.241111}
  {\bibfield  {journal} {\bibinfo  {journal} {Phys. Rev.}\ }\textbf {\bibinfo
  {volume} {B94}},\ \bibinfo {pages} {241111} (\bibinfo {year} {2016})},\
  \Eprint {http://arxiv.org/abs/1603.08376} {arXiv:1603.08376} \BibitemShut
  {NoStop}%
\bibitem [{\citenamefont {Ayyar}\ and\ \citenamefont
  {Chandrasekharan}(2016{\natexlab{b}})}]{Ayyar:2016lxq}%
  \BibitemOpen
  \bibfield  {author} {\bibinfo {author} {\bibfnamefont {V.}~\bibnamefont
  {Ayyar}}\ and\ \bibinfo {author} {\bibfnamefont {S.}~\bibnamefont
  {Chandrasekharan}},\ }\bibfield  {title} {\enquote {\bibinfo {title}
  {{Fermion masses through four-fermion condensates}},}\ }\href {\doibase
  10.1007/JHEP10(2016)058} {\bibfield  {journal} {\bibinfo  {journal} {JHEP}\
  }\textbf {\bibinfo {volume} {10}},\ \bibinfo {pages} {058} (\bibinfo {year}
  {2016}{\natexlab{b}})},\ \Eprint {http://arxiv.org/abs/1606.06312}
  {arXiv:1606.06312} \BibitemShut {NoStop}%
\bibitem [{\citenamefont {Ayyar}(2016)}]{Ayyar:2016nqh}%
  \BibitemOpen
  \bibfield  {author} {\bibinfo {author} {\bibfnamefont {V.}~\bibnamefont
  {Ayyar}},\ }\bibfield  {title} {\enquote {\bibinfo {title} {{Search for a
  continuum limit of the PMS phase}},}\ }\href@noop {} {\bibfield  {journal}
  {\bibinfo  {journal} {PoS}\ }\textbf {\bibinfo {volume} {LATTICE2016}},\
  \bibinfo {pages} {327} (\bibinfo {year} {2016})},\ \Eprint
  {http://arxiv.org/abs/1611.00280} {arXiv:1611.00280} \BibitemShut {NoStop}%
\bibitem [{\citenamefont {Catterall}\ and\ \citenamefont
  {Schaich}(2017)}]{Catterall:2016dzf}%
  \BibitemOpen
  \bibfield  {author} {\bibinfo {author} {\bibfnamefont {S.}~\bibnamefont
  {Catterall}}\ and\ \bibinfo {author} {\bibfnamefont {D.}~\bibnamefont
  {Schaich}},\ }\bibfield  {title} {\enquote {\bibinfo {title} {{Novel phases
  in strongly coupled four-fermion theories}},}\ }\href {\doibase
  10.1103/PhysRevD.96.034506} {\bibfield  {journal} {\bibinfo  {journal} {Phys.
  Rev.}\ }\textbf {\bibinfo {volume} {D96}},\ \bibinfo {pages} {034506}
  (\bibinfo {year} {2017})},\ \Eprint {http://arxiv.org/abs/1609.08541}
  {arXiv:1609.08541} \BibitemShut {NoStop}%
\bibitem [{\citenamefont {Schaich}\ and\ \citenamefont
  {Catterall}(2018)}]{Schaich:2017czc}%
  \BibitemOpen
  \bibfield  {author} {\bibinfo {author} {\bibfnamefont {D.}~\bibnamefont
  {Schaich}}\ and\ \bibinfo {author} {\bibfnamefont {S.}~\bibnamefont
  {Catterall}},\ }\bibfield  {title} {\enquote {\bibinfo {title} {{Phases of a
  strongly coupled four-fermion theory}},}\ }\href {\doibase
  10.1051/epjconf/201817503004} {\bibfield  {journal} {\bibinfo  {journal} {EPJ
  Web Conf.}\ }\textbf {\bibinfo {volume} {175}},\ \bibinfo {pages} {03004}
  (\bibinfo {year} {2018})},\ \Eprint {http://arxiv.org/abs/1710.08137}
  {arXiv:1710.08137} \BibitemShut {NoStop}%
\bibitem [{\citenamefont {Bock}\ \emph {et~al.}(1992)\citenamefont {Bock},
  \citenamefont {Smit},\ and\ \citenamefont {Vink}}]{Bock:1992yr}%
  \BibitemOpen
  \bibfield  {author} {\bibinfo {author} {\bibfnamefont {W.}~\bibnamefont
  {Bock}}, \bibinfo {author} {\bibfnamefont {J.}~\bibnamefont {Smit}}, \ and\
  \bibinfo {author} {\bibfnamefont {J.~C.}\ \bibnamefont {Vink}},\ }\bibfield
  {title} {\enquote {\bibinfo {title} {{Fermion Higgs model with reduced
  staggered fermions}},}\ }\href {\doibase 10.1016/0370-2693(92)91049-F}
  {\bibfield  {journal} {\bibinfo  {journal} {Phys. Lett.}\ }\textbf {\bibinfo
  {volume} {B291}},\ \bibinfo {pages} {297--305} (\bibinfo {year} {1992})},\
  \Eprint {http://arxiv.org/abs/hep-lat/9206008} {hep-lat/9206008} \BibitemShut
  {NoStop}%
\bibitem [{\citenamefont {Catterall}\ and\ \citenamefont
  {Butt}(2018)}]{Catterall:2017ogi}%
  \BibitemOpen
  \bibfield  {author} {\bibinfo {author} {\bibfnamefont {S.}~\bibnamefont
  {Catterall}}\ and\ \bibinfo {author} {\bibfnamefont {N.}~\bibnamefont
  {Butt}},\ }\bibfield  {title} {\enquote {\bibinfo {title} {{Topology and
  strong four fermion interactions in four dimensions}},}\ }\href {\doibase
  10.1103/PhysRevD.97.094502} {\bibfield  {journal} {\bibinfo  {journal} {Phys.
  Rev.}\ }\textbf {\bibinfo {volume} {D97}},\ \bibinfo {pages} {094502}
  (\bibinfo {year} {2018})},\ \Eprint {http://arxiv.org/abs/1708.06715}
  {arXiv:1708.06715} \BibitemShut {NoStop}%
\bibitem [{\citenamefont {Fidkowski}\ and\ \citenamefont
  {Kitaev}(2010)}]{Fidkowski:2009dba}%
  \BibitemOpen
  \bibfield  {author} {\bibinfo {author} {\bibfnamefont {L.}~\bibnamefont
  {Fidkowski}}\ and\ \bibinfo {author} {\bibfnamefont {A.}~\bibnamefont
  {Kitaev}},\ }\bibfield  {title} {\enquote {\bibinfo {title} {{The effects of
  interactions on the topological classification of free fermion systems}},}\
  }\href {\doibase 10.1103/PhysRevB.81.134509} {\bibfield  {journal} {\bibinfo
  {journal} {Phys. Rev.}\ }\textbf {\bibinfo {volume} {B81}},\ \bibinfo {pages}
  {134509} (\bibinfo {year} {2010})},\ \Eprint {http://arxiv.org/abs/0904.2197}
  {arXiv:0904.2197} \BibitemShut {NoStop}%
\bibitem [{\citenamefont {Morimoto}\ \emph {et~al.}(2015)\citenamefont
  {Morimoto}, \citenamefont {Furusaki},\ and\ \citenamefont
  {Mudry}}]{Morimoto:2015lua}%
  \BibitemOpen
  \bibfield  {author} {\bibinfo {author} {\bibfnamefont {T.}~\bibnamefont
  {Morimoto}}, \bibinfo {author} {\bibfnamefont {A.}~\bibnamefont {Furusaki}},
  \ and\ \bibinfo {author} {\bibfnamefont {C.}~\bibnamefont {Mudry}},\
  }\bibfield  {title} {\enquote {\bibinfo {title} {{Breakdown of the
  topological classification $\mathbb{Z}$ for gapped phases of noninteracting
  fermions by quartic interactions}},}\ }\href {\doibase
  10.1103/PhysRevB.92.125104} {\bibfield  {journal} {\bibinfo  {journal} {Phys.
  Rev.}\ }\textbf {\bibinfo {volume} {B92}},\ \bibinfo {pages} {125104}
  (\bibinfo {year} {2015})},\ \Eprint {http://arxiv.org/abs/1505.06341}
  {arXiv:1505.06341} \BibitemShut {NoStop}%
\bibitem [{\citenamefont {Stephenson}\ and\ \citenamefont
  {Thornton}(1988)}]{Stephenson:1988td}%
  \BibitemOpen
  \bibfield  {author} {\bibinfo {author} {\bibfnamefont {D.}~\bibnamefont
  {Stephenson}}\ and\ \bibinfo {author} {\bibfnamefont {A.}~\bibnamefont
  {Thornton}},\ }\bibfield  {title} {\enquote {\bibinfo {title}
  {{Nonperturbative Yukawa Couplings}},}\ }\href {\doibase
  10.1016/0370-2693(88)91800-X} {\bibfield  {journal} {\bibinfo  {journal}
  {Phys. Lett.}\ }\textbf {\bibinfo {volume} {B212}},\ \bibinfo {pages}
  {479--482} (\bibinfo {year} {1988})}\BibitemShut {NoStop}%
\bibitem [{\citenamefont {Hasenfratz}\ and\ \citenamefont
  {Neuhaus}(1989)}]{Hasenfratz:1988vc}%
  \BibitemOpen
  \bibfield  {author} {\bibinfo {author} {\bibfnamefont {A.}~\bibnamefont
  {Hasenfratz}}\ and\ \bibinfo {author} {\bibfnamefont {T.}~\bibnamefont
  {Neuhaus}},\ }\bibfield  {title} {\enquote {\bibinfo {title}
  {{Nonperturbative Study of the Strongly Coupled Scalar Fermion Model}},}\
  }\href {\doibase 10.1016/0370-2693(89)90899-X} {\bibfield  {journal}
  {\bibinfo  {journal} {Phys. Lett.}\ }\textbf {\bibinfo {volume} {B220}},\
  \bibinfo {pages} {435--440} (\bibinfo {year} {1989})}\BibitemShut {NoStop}%
\bibitem [{\citenamefont {Lee}\ \emph {et~al.}(1990{\natexlab{a}})\citenamefont
  {Lee}, \citenamefont {Shigemitsu},\ and\ \citenamefont
  {Shrock}}]{Lee:1989xq}%
  \BibitemOpen
  \bibfield  {author} {\bibinfo {author} {\bibfnamefont {I.-H.}\ \bibnamefont
  {Lee}}, \bibinfo {author} {\bibfnamefont {J.}~\bibnamefont {Shigemitsu}}, \
  and\ \bibinfo {author} {\bibfnamefont {R.~E.}\ \bibnamefont {Shrock}},\
  }\bibfield  {title} {\enquote {\bibinfo {title} {{Lattice Study of a Yukawa
  Theory With a Real Scalar Field}},}\ }\href {\doibase
  10.1016/0550-3213(90)90309-2} {\bibfield  {journal} {\bibinfo  {journal}
  {Nucl. Phys.}\ }\textbf {\bibinfo {volume} {B330}},\ \bibinfo {pages}
  {225--260} (\bibinfo {year} {1990}{\natexlab{a}})}\BibitemShut {NoStop}%
\bibitem [{\citenamefont {Lee}\ \emph {et~al.}(1990{\natexlab{b}})\citenamefont
  {Lee}, \citenamefont {Shigemitsu},\ and\ \citenamefont
  {Shrock}}]{Lee:1989mi}%
  \BibitemOpen
  \bibfield  {author} {\bibinfo {author} {\bibfnamefont {I.-H.}\ \bibnamefont
  {Lee}}, \bibinfo {author} {\bibfnamefont {J.}~\bibnamefont {Shigemitsu}}, \
  and\ \bibinfo {author} {\bibfnamefont {R.~E.}\ \bibnamefont {Shrock}},\
  }\bibfield  {title} {\enquote {\bibinfo {title} {{Study of Different Lattice
  Formulations of a Yukawa Model With a Real Scalar Field}},}\ }\href {\doibase
  10.1016/0550-3213(90)90664-Y} {\bibfield  {journal} {\bibinfo  {journal}
  {Nucl. Phys.}\ }\textbf {\bibinfo {volume} {B334}},\ \bibinfo {pages}
  {265--278} (\bibinfo {year} {1990}{\natexlab{b}})}\BibitemShut {NoStop}%
\bibitem [{\citenamefont {Bock}\ and\ \citenamefont {De}(1990)}]{Bock:1990cx}%
  \BibitemOpen
  \bibfield  {author} {\bibinfo {author} {\bibfnamefont {W.}~\bibnamefont
  {Bock}}\ and\ \bibinfo {author} {\bibfnamefont {A.~K.}\ \bibnamefont {De}},\
  }\bibfield  {title} {\enquote {\bibinfo {title} {{Unquenched Investigation of
  Fermion Masses in a Chiral Fermion Theory on the Lattice}},}\ }\href
  {\doibase 10.1016/0370-2693(90)90135-S} {\bibfield  {journal} {\bibinfo
  {journal} {Phys. Lett.}\ }\textbf {\bibinfo {volume} {B245}},\ \bibinfo
  {pages} {207--212} (\bibinfo {year} {1990})}\BibitemShut {NoStop}%
\bibitem [{\citenamefont {Abada}\ and\ \citenamefont
  {Shrock}(1991)}]{Abada:1990ds}%
  \BibitemOpen
  \bibfield  {author} {\bibinfo {author} {\bibfnamefont {A.}~\bibnamefont
  {Abada}}\ and\ \bibinfo {author} {\bibfnamefont {R.~E.}\ \bibnamefont
  {Shrock}},\ }\bibfield  {title} {\enquote {\bibinfo {title} {{Results from a
  strong coupling expansion for a lattice Yukawa model with a real scalar
  field}},}\ }\href {\doibase 10.1103/PhysRevD.43.R304} {\bibfield  {journal}
  {\bibinfo  {journal} {Phys. Rev.}\ }\textbf {\bibinfo {volume} {D43}},\
  \bibinfo {pages} {304--307} (\bibinfo {year} {1991})}\BibitemShut {NoStop}%
\bibitem [{\citenamefont {Hasenfratz}\ \emph {et~al.}(1991)\citenamefont
  {Hasenfratz}, \citenamefont {Hasenfratz}, \citenamefont {Jansen},
  \citenamefont {Kuti},\ and\ \citenamefont {Shen}}]{Hasenfratz:1991it}%
  \BibitemOpen
  \bibfield  {author} {\bibinfo {author} {\bibfnamefont {A.}~\bibnamefont
  {Hasenfratz}}, \bibinfo {author} {\bibfnamefont {P.}~\bibnamefont
  {Hasenfratz}}, \bibinfo {author} {\bibfnamefont {K.}~\bibnamefont {Jansen}},
  \bibinfo {author} {\bibfnamefont {J.}~\bibnamefont {Kuti}}, \ and\ \bibinfo
  {author} {\bibfnamefont {Y.}~\bibnamefont {Shen}},\ }\bibfield  {title}
  {\enquote {\bibinfo {title} {{The Equivalence of the top quark condensate and
  the elementary Higgs field}},}\ }\href {\doibase
  10.1016/0550-3213(91)90607-Y} {\bibfield  {journal} {\bibinfo  {journal}
  {Nucl. Phys.}\ }\textbf {\bibinfo {volume} {B365}},\ \bibinfo {pages}
  {79--97} (\bibinfo {year} {1991})}\BibitemShut {NoStop}%
\bibitem [{\citenamefont {Gerhold}\ and\ \citenamefont
  {Jansen}(2007{\natexlab{a}})}]{Gerhold:2007yb}%
  \BibitemOpen
  \bibfield  {author} {\bibinfo {author} {\bibfnamefont {P.}~\bibnamefont
  {Gerhold}}\ and\ \bibinfo {author} {\bibfnamefont {K.}~\bibnamefont
  {Jansen}},\ }\bibfield  {title} {\enquote {\bibinfo {title} {{The Phase
  structure of a chirally invariant lattice Higgs-Yukawa model for small and
  for large values of the Yukawa coupling constant}},}\ }\href {\doibase
  10.1088/1126-6708/2007/09/041} {\bibfield  {journal} {\bibinfo  {journal}
  {JHEP}\ }\textbf {\bibinfo {volume} {0709}},\ \bibinfo {pages} {041}
  (\bibinfo {year} {2007}{\natexlab{a}})},\ \Eprint
  {http://arxiv.org/abs/0705.2539} {arXiv:0705.2539} \BibitemShut {NoStop}%
\bibitem [{\citenamefont {Gerhold}\ and\ \citenamefont
  {Jansen}(2007{\natexlab{b}})}]{Gerhold:2007gx}%
  \BibitemOpen
  \bibfield  {author} {\bibinfo {author} {\bibfnamefont {P.}~\bibnamefont
  {Gerhold}}\ and\ \bibinfo {author} {\bibfnamefont {K.}~\bibnamefont
  {Jansen}},\ }\bibfield  {title} {\enquote {\bibinfo {title} {{The Phase
  structure of a chirally invariant lattice Higgs-Yukawa model - numerical
  simulations}},}\ }\href {\doibase 10.1088/1126-6708/2007/10/001} {\bibfield
  {journal} {\bibinfo  {journal} {JHEP}\ }\textbf {\bibinfo {volume} {0710}},\
  \bibinfo {pages} {001} (\bibinfo {year} {2007}{\natexlab{b}})},\ \Eprint
  {http://arxiv.org/abs/0707.3849} {arXiv:0707.3849} \BibitemShut {NoStop}%
\end{thebibliography}%

\end{document}